\documentclass[aps,superscriptaddress,twocolumn,noshowkeys,noshowpacs,nofootinbib]{revtex4-2}
\usepackage{epsf,amsmath,graphicx,amssymb,url,hyperref,dcolumn}
\usepackage{mathtools}
\usepackage{tabu}
\usepackage{color}
\usepackage[normalem]{ulem}
\usepackage{adjustbox}
\usepackage{array}
\usepackage{xcolor}
\usepackage[justification=raggedright,singlelinecheck=true]{caption}
\usepackage{orcidlink}

\usepackage{subcaption}
\begin{document}

\title{ Partially Thermalized eV-Scale Sterile Neutrinos in Dynamical Dark Energy and Non-Flat Cosmologies}

\author{Gopal Kashyap\,\orcidlink{0000-0001-9001-4905}}
\email{gopal.hari@vit.ac.in}
\affiliation{Department of Physics, School of Advanced Sciences, Vellore Institute of Technology, Vellore 632014, Tamil Nadu, India}
\author{Kanhaiya Lal Pandey\,\orcidlink{0000-0003-3536-1730}}
\email{kanhaiyalal.pandey@vit.ac.in}
\affiliation{Department of Physics, School of Advanced Sciences, Vellore Institute of Technology, Vellore 632014, Tamil Nadu, India}
\date{\today}

\begin{abstract}
    
We investigate cosmological constraints on a $(3+1)$ neutrino framework containing an additional partially thermalized, eV-scale sterile neutrino
alongside the three active species. Rather than assuming full thermalization, we allow the sterile abundance to vary independently
through its physical mass $m_s$ and contribution $\Delta N_{\rm eff}$ to the effective number of relativistic species. The active-neutrino sector
is analyzed separately for normal and inverted mass ordering, while the sterile state is modeled as a thermally distributed non-cold relic with
a temperature below that of the active-neutrino background. We consider flat and non-flat $\Lambda$CDM and $w_0w_a$CDM cosmologies, all including
the same $(3+1)\nu$ sector. The models are constrained using Planck 2018 CMB data, DESI DR2 BAO, Pantheon+ Type Ia supernovae, a conservative compilation of
$f\sigma_8$ measurements, and the recent Hubble Distance Network constraint, $H_0=73.50\pm0.81~{\rm km\,s^{-1}\,Mpc^{-1}}$. Oscillation-informed priors are imposed on the active-neutrino mass splittings, while the eV scale sterile-mass range is motivated by short-baseline oscillation searches and global $(3+1)$ analyses. We find that allowing additional freedom in the late-time cosmological background substantially changes the inferred sterile-neutrino sector.
The effective sterile mass increases from $m_{s,\rm eff}\sim0.11\pm0.05~{\rm eV}$ in flat $\Lambda$CDM$+(3+1)\nu$ to $\sim0.35\pm0.12~{\rm eV}$ in the non-flat $w_0w_a$CDM$+(3+1)\nu$ model. The corresponding sterile abundance shifts from $\Delta N_{\rm eff}\lesssim0.1$ to $\Delta N_{\rm eff}\simeq0.35\pm0.15$, while remaining well below the fully thermalized limit, $\Delta N_{\rm eff}\simeq1$. A direct comparison with the induced prior on $m_{s,\rm eff}$ shows that the posterior displacement toward larger effective sterile masses is not solely a consequence of transformed prior volume. The non-flat $w_0w_a$CDM models also show $\sim 2.8\sigma$ preference for open spatial geometry, $\Omega_k\simeq0.0045\pm0.0016$, and yield $H_0\simeq70.9\pm0.7~{\rm km\,s^{-1}\,Mpc^{-1}}$ and $S_8\simeq0.785\pm0.015$. These models improve the minimum $\chi^2$ by approximately $21$-$25$ relative to flat
$\Lambda$CDM$+(3+1)\nu$. Our results therefore show that the cosmological viability of a partially thermalized eV-scale sterile neutrino is strongly correlated with assumptions about spatial curvature and the late-time dark-energy sector.
\end{abstract}

\maketitle

\section{Introduction}
Neutrino oscillation experiments have established that neutrinos are massive and that the three active flavor states are mixtures of mass eigenstates, providing one of the clearest indications of physics beyond the minimal Standard Model \cite{1998PhRvL..81.1562F, 2002PhRvL..89a1301A, 2022PTEP.2022h3C01W}. The possibility of additional sterile neutrino states, which do not participate directly in standard weak interactions, has been extensively investigated in connection with anomalies reported by short-baseline (SBL) experiments, including LSND and MiniBooNE \cite{2001PhRvD..64k2007A, 2018PhRvL.121v1801A}. Reactor, accelerator, and other disappearance measurements, such as, Daya Bay, MINOS/MINOS+, and IceCube have subsequently provided important constraints on the corresponding active–sterile mixing parameters, while also highlighting persistent tensions among different experimental datasets \cite{2015JPhG...43c3001G, 2018JHEP...08..010D, 2018PhLB..787...56A}. 
Neutrinos provide an important link between particle physics and cosmology, with their masses and relativistic energy density leaving characteristic imprints on the expansion history, cosmic microwave background (CMB), and large-scale structure \cite{2006PhR...429..307L, 2006ARNPS..56..137H, 2015PhRvD..92l3535A, 2024arXiv240419322D}. 
A sterile neutrino with a mass at the eV scale is particularly interesting cosmologically because, if sufficiently populated in the early Universe, it contributes to the relativistic energy density around recombination and later behaves as a component of hot dark matter. Cosmological observations therefore provide a complementary probe of light sterile-neutrino scenarios, sensitive to both their abundance and their impact on the expansion history and growth of cosmic structure.

The cosmological implications of sterile neutrinos depend crucially on their production history and phase-space distribution. In the simplest scenario, a fully thermalized sterile neutrino contributes approximately one additional effective neutrino species, $\Delta N_{\rm eff} \simeq 1$, while its non-relativistic energy density is determined by its physical mass. Such a scenario is strongly constrained by combinations of cosmic microwave background (CMB) and large-scale-structure observations \cite{2020A&A...641A...6P, 2012JCAP...07..025H, 2015PhRvD..91f5021A}. In particular, the Planck 2018 analysis found stringent constraints on both the effective number of relativistic species and the effective mass of an additional sterile relic, strongly disfavoring a fully thermalized eV-scale sterile neutrino when combined with baryon acoustic oscillation (BAO) data \cite{2020A&A...641A...6P}. These constraints, however, should not be interpreted as model-independent exclusions of sterile neutrinos, since the cosmological abundance of a sterile state need not correspond to complete thermalization. Mechanisms such as suppressed production, non-standard neutrino interactions, or a modified thermal history can lead to a reduced sterile abundance and consequently to $\Delta N_{\rm eff} < 1$ \cite{2012JCAP...07..025H, 2015PhRvD..91f5021A, 2015PhRvD..92l3534D, 2006PhR...429..307L}.

An important recent development in this direction is the systematic comparison of different sterile-neutrino production scenarios. Ladeira et al. \cite{2026PhRvD.114d3511L} considered fully thermalized, different-temperature thermal relic, and Dodelson–Widrow-like realizations and showed that current cosmological data place substantially stronger constraints on the effective sterile abundance than on its physical mass when the sterile population is suppressed. In particular, their different-temperature sterile scenario remains considerably less constrained than the fully thermalized case, emphasizing the importance of distinguishing the physical sterile mass from its cosmological abundance. This result is consistent with the broader picture in which cosmological observations constrain the energy density and free-streaming effects of relic neutrinos rather than directly measuring their microscopic production mechanism. The distinction between physical and effective sterile-neutrino masses has also been incorporated in earlier Planck analyses of massive sterile relics \cite{2020A&A...641A...6P}. Thus, a meaningful assessment of eV-scale sterile neutrinos requires the thermalization level and the physical mass to be treated as separate quantities.

The interpretation of neutrino cosmology is further complicated by degeneracies with the late-time expansion history. Massive neutrinos alter the background evolution and suppress the growth of matter perturbations through free streaming, while changes in the dark-energy equation of state can partially compensate these effects in cosmological observables \cite{2012arXiv1212.6154L, 2012PhRvD..85l3521S, 2017JCAP...02..052A, 2018PhRvD..97l3504P, 2022ApJ...934..113S, 2022PhRvD.106d3540D, 2023PhRvD.108h3509D}. 
More recent analyses using DESI BAO data have likewise emphasized the sensitivity of neutrino-mass constraints to assumptions about the dark-energy sector, with evolving dark energy providing additional freedom in accommodating the cosmological neutrino-mass scale \cite{2026NewAR.10301759G, 2026PhRvD.114d3543P, 2026PhRvD.113l3506C, 2026PhRvD.113h3503L}. This issue is particularly relevant in light of the DESI BAO measurements, which have provided indications of departures from a cosmological constant when combined with complementary distance measurements \cite{2025PhRvD.111b3532L, 2025PhRvD.112h3513E, 2025JCAP...02..021A}. Consequently, sterile-neutrino constraints obtained within $\Lambda$CDM may not fully characterize the allowed parameter space when the expansion history is generalized \cite{2013JCAP...11..018D, 2026PhRvD.113h3503L}.

The increasing precision of late-time cosmological observations makes it possible to investigate these effects using complementary probes. The Planck CMB measurements tightly constrain the early-Universe radiation content and neutrino-induced changes to the CMB anisotropies \cite{2020A&A...641A...6P}, while BAO measurements provide precise information about the distance-redshift relation and the expansion history. The recent DESI DR2 BAO measurements have substantially improved the precision of cosmological constraints and have also highlighted the interplay between neutrino masses and dynamical dark energy \cite{2025JCAP...02..021A}. Type Ia supernovae provide an independent probe of the late-time expansion history, while redshift-space-distortion measurements constrain the growth of structure through $f\sigma_8$. Combining these probes is therefore particularly useful for separating the effects of sterile-neutrino abundance, neutrino free streaming, dark-energy dynamics, and spatial curvature. In addition, local measurements of the Hubble constant provide a complementary constraint on the late-time expansion rate and can alter the allowed degeneracies between $N_{\rm eff}$, neutrino masses, and dark energy.

In this work, we investigate a (3 + 1) neutrino framework containing three active neutrinos and an additional partially thermalized, eV-scale sterile neutrino. The sterile sector is characterized independently by its physical mass and its contribution $\Delta N_{\rm eff}$ to the effective number of relativistic species, and is modeled as a thermal non-cold relic with a temperature lower than that of the active-neutrino background. This allows us to explore the sterile-neutrino parameter space continuously between strongly suppressed and fully populated regimes while consistently accounting for the transition of the sterile state from relativistic to non-relativistic behavior. We consider both normal and inverted ordering (hereafter NO and IO) of the active-neutrino masses and investigate a sequence of cosmological models ranging from flat $\Lambda$CDM to flat $w_0 w_a$CDM and spatially curved $w_0 w_a$CDM, the latter two models are studied independently for NO and IO both. Our analysis combines Planck 2018 CMB data, DESI DR2 BAO measurements, Pantheon+ Type Ia supernovae, conservative $f\sigma_8$ measurements, and a Gaussian $H_0$ prior from the recent Hubble Distance Network measurement. We additionally incorporate neutrino-oscillation adapted from recent JUNO \cite{2026Natur.654..343J}, 
and earlier short-baseline (SBL) experiments \cite{2017JHEP...06..135G}. Within this framework, we quantify the joint constraints on the sterile-neutrino mass and thermalization level and examine their degeneracies with the active-neutrino mass sector, dark-energy dynamics, and spatial curvature. This enables us to assess whether the cosmological viability of an eV-scale sterile neutrino remains substantially altered once partial thermalization and extensions of the late-time cosmological model are simultaneously taken into account.

The paper is organized as follows. Section~\ref{sec:model} introduces the cosmological and neutrino framework, including the active-neutrino mass ordering, the sterile-neutrino sector, the background cosmology, and the growth of structure. Section~\ref{sec:data} describes the observational data sets used in the analysis. The statistical methodology, including the parameter priors,
numerical implementation, and model-comparison criteria, is presented in
Sec.~\ref{sec:stat}. The resulting cosmological constraints are discussed
in Sec.~\ref{sec:results}, and our main conclusions are summarized in
Sec.~\ref{sec:conclusions}.

\section{Cosmological and Neutrino Framework}
\label{sec:model}
\subsection{Active-neutrino masses and ordering}
\label{subsec_act_nu}
We parameterize the active-neutrino sector through the lightest mass $m_1\equiv m_{\rm lightest}$ for the normal ordering, and,
\begin{equation}
m_2 = \sqrt{m_1^2+\Delta m_{21}^2}, \qquad m_3 = \sqrt{m_1^2+\Delta m_{31}^2},
\end{equation}
where $\Delta m_{31}^2>0$.
For the inverted ordering, $m_3\equiv m_{\rm lightest}$ with the NuFIT convention 
in which $\Delta m_{32}^2=m_3^2-m_2^2<0$, and,
\begin{align}
m_2 = \sqrt{m_3^2+|\Delta m_{32}^2|}, \qquad m_1 = \sqrt{m_2^2-\Delta m_{21}^2}.
\end{align}

The active-neutrino mass sum is
\begin{equation}
\sum m_{\nu,\rm active}
\equiv
m_1+m_2+m_3.
\label{eq:active_sum}
\end{equation}
The different ordering-dependent mass floors in
Eq.~(\ref{eq:active_sum}) provide the principal direct cosmological
distinction between NO and IO.

We implement all three active neutrinos as massive species, in the publicly available \texttt{CLASS} (Cosmic Linear Anisotropy Solving System) Boltzmann code \cite{CLASS}.
For the three active species we use $T_{\rm active}/T_\gamma=0.71611$, consistent with a more accurate treatment of non-instantaneous neutrino decoupling and finite-temperature QED corrections \cite{2021JCAP...04..073B}. A residual ultra-relativistic contribution, $N_{\rm ur}=0.00441$ is included to recover the standard value of $N^{\rm SM}_{\rm eff}\simeq3.044$ for the active-neutrino sector. 

\subsection{The eV-scale sterile neutrino}
\label{subsec_str_nu}

The fourth neutrino state is modeled as a thermally distributed non-cold relic. Unlike the fully thermalized limit, in which the sterile state attains the same thermal abundance and temperature as the active neutrino background, we allow for incomplete thermalization of the sterile species.  The sterile sector is therefore characterized by a suppressed contribution to the relativistic energy density, parameterized by $(\Delta N_{\rm eff})$, and by its physical mass ($m_s$).
The total effective number of relativistic species can be written as
\begin{equation}
N_{\rm eff}=N_{\rm eff}^{\rm SM}+\Delta N_{\rm eff}, \quad \text{where}, \quad N_{\rm eff}^{\rm SM}=3.044
\end{equation}
denotes the Standard Model prediction for the three active neutrinos. Thus, in the partially thermalized scenario, $0<\Delta N_{\rm eff}<1,$
with the limiting case $(\Delta N_{\rm eff}=1)$ corresponding to a fully thermalized neutrino species.
For a thermally distributed sterile neutrino with temperature $T_s$, its relativistic energy density is proportional to $T_s^4$. Hence, its contribution to the radiation density is related to its temperature by,
\begin{equation}
\frac{T_s}{T_{\nu,{\rm active}}}=(\Delta N_{\rm eff})^{1/4},\label{eq:Ts} \quad {\rm where,} \quad \frac{T_{\nu,{\rm active}}}{T_\gamma} \simeq \left(\frac{4}{11}\right)^{1/3}.
\end{equation}

We assume a Fermi–Dirac phase-space distribution for the sterile neutrino. Since the number density scales as $T_s^3$, partial thermalization also reduces the late-time abundance of the sterile species. In the non-relativistic regime, its cosmological density can therefore be expressed in terms of an effective mass,
\begin{equation}
m_{s,\rm eff}=m_s(\Delta N_{\rm eff})^{3/4}.
\label{eq:m4eff}
\end{equation}

Thus, for a given physical mass $m_s$, the cosmological impact of the sterile neutrino is suppressed when $\Delta N_{\rm eff}$ is below unity. 

\subsection{The background cosmology}

We consider a Friedmann-Lemaitre-Robertson-Walker background with
matter, radiation, dark energy, and spatial curvature.
The Friedmann equation for this model can be written as,
\begin{multline}
\frac{H^2(a)}{H_0^2}=\Omega_\gamma a^{-4}+(\Omega_{\rm c}+\Omega_{\rm b})
a^{-3} +\Omega_k a^{-2} \\
+\Omega_{\rm de}f_{\rm de}(a) + \Omega_\nu(a),\label{eq:Ez}
\end{multline}
where $\Omega_\gamma$, $\Omega_c$, $\Omega_b$, $\Omega_k$, $\Omega_{\rm{de}}$, and $\Omega_{\nu,0} \equiv \Omega_\nu(a=1)$ denote the present-day density parameters for photons, cold dark matter, baryons, curvature, dark energy, and total neutrinos (active and sterile), respectively, satisfying $\Omega_\gamma + \Omega_c + \Omega_b + \Omega_k + \Omega_{\rm{de}} + \Omega_{\nu,0} = 1$.
For dynamical dark energy we adopt the
Chevallier-Polarski-Linder (CPL) form \cite{Chevallier2001,Linder2003},
\begin{equation}
w(a)=w_0+w_a(1-a),
\end{equation}
The corresponding dark-energy density evolves as
\begin{equation}
f_{\rm de}(a)=a^{-3(1+w_0+w_a)}
\exp\left[-{3w_a (1-a)}\right].
\label{eq:CPLrho}
\end{equation}

The spatial-curvature parameter is defined using the standard
convention
\begin{equation}
\Omega_k=-\frac{k}{a_0^2H_0^2}.
\end{equation}
Thus $\Omega_k>0$ corresponds to an open universe, while
$\Omega_k<0$ corresponds to a closed universe.

For the neutrino sector, the total neutrino energy density contains contributions from all three active neutrinos and the partially thermalized sterile species. In our analysis, all four neutrino species are treated explicitly as non-cold relics, 
allowing their effects on both the background expansion and the evolution of cosmological perturbations to be consistently included.
The combined energy density parameter can be given as,
$$\Omega_\nu(a) = \frac{\sum \rho_{\nu,\rm{active}}(a) + \rho_s(a)}{\rho_{\rm{c},0}}.$$

The total neutrino energy density does not obey a single power-law scaling throughout cosmic evolution. At sufficiently high redshift, when the neutrino species are relativistic, their energy densities scale approximately as $(1+z)^4$, contributing to the radiation sector. As the Universe expands and individual massive neutrino species become non-relativistic, their energy densities gradually approach the matter-like scaling $(1+z)^3$. Since the active and sterile neutrinos can have different masses and temperature distributions, their relativistic-to-non-relativistic transitions occur at different redshifts.

The redshift evolution of the total neutrino energy density is calculated directly from the corresponding phase-space distributions, with the evolution implemented by adapting the  \texttt{CLASS} code, as detailed in \ref{subsec_act_nu} and \ref{subsec_str_nu}.

\subsection{Growth of structure}
\label{sec_fs8}

Massive neutrinos suppress clustering on scales smaller than their free-streaming length. The observable commonly reported by redshift-space-distortion surveys is $f\sigma_8(z)$, defined as the product of the growth rate $f(z)$ and the rms matter 
fluctuation amplitude $\sigma_8(z)$. In terms of the scale factor $a$, this quantity is computed as, 
\begin{equation}
f\sigma_8(a) = a \frac{\delta_m'(a)}{\delta_m(1)} \sigma_{8,0},
\end{equation}
where
\begin{equation}
f(a) = \frac{d\ln D(a)}{d\ln a}
\end{equation}
is the logarithmic linear growth rate, $D(a)$ denotes the linear growth factor, and the prime denotes differentiation with respect to 
the scale factor $a$. 

Rather than imposing a scale-independent analytic growth approximation, we evaluate the corresponding quantity directly from
\texttt{CLASS} for every sampled cosmology. This is particularly important in the presence of massive active and sterile neutrinos.

\section{Observational data}
\label{sec:data}
\subsection{Neutrino-oscillation data}
\label{Nu_Osc_data}
Recent results from the Jiangmen Underground Neutrino Observatory (JUNO) \cite{2026Natur.654..343J} have opened a new era of precision neutrino-oscillation measurements. With its long-baseline reactor configuration and high energy resolution, JUNO is designed to measure the neutrino mass-squared splittings and mixing parameters with unprecedented precision, while providing an important avenue toward determining the neutrino mass ordering. In our analysis, we incorporate this improved oscillation information through priors on the active-neutrino mass-squared splittings motivated by the recent JUNO measurements.

\begin{table}[ht]
\centering
\caption{Neutrino mass-squared differences adopted in our analysis.}
\label{tab:mass_splittings}
\begin{tabular}{lcc}
\hline
\hline
Parameter & Value & Mass ordering \\
\hline
$\Delta m_{21}^2$ & $7.49\times10^{-5}\ {\rm eV}^2$ & - \\
$\Delta m_{31}^2$ & $2.513\times10^{-3}\ {\rm eV}^2$ & NO \\
$\Delta m_{32}^2$ & $-2.483\times10^{-3}\ {\rm eV}^2$ & IO \\
\hline
\end{tabular}
\end{table}
The precise values of these splittings have a negligible impact on the
cosmological constraints compared with the uncertainty in the absolute
mass scale.

For the sterile-neutrino sector, we adopt priors motivated by short-baseline (SBL) oscillation phenomenology within the (3+1) framework. In the (3+1) framework, the relevant mass-squared splitting is $\Delta m_{41}^{2}=m_4^2-m_1^2,$ where $m_4\equiv m_s$ denotes the sterile mass eigenstate. Since the active-neutrino masses considered in this work are sub-eV, $m_1^2\ll m_s^2$ over the parameter region of interest, and hence $m_s\simeq\sqrt{\Delta m_{41}^{2}}$. The global analysis of Gariazzo et al. \cite{2017JHEP...06..135G} finds an allowed eV-scale region in the mass-squared splitting $\Delta m_{41}^{2}$, with a best-fit value near $1.7~{\rm eV}^{2}$ and an allowed region extending to approximately $1.3~{\rm eV}^{2}$ at $2\sigma$. We therefore restrict the physical sterile-neutrino mass to $0.5\leq {m_s}/{\rm eV}\leq1.5$, thereby focusing on the eV-scale region suggested by SBL oscillation data rather than allowing an unrestricted sterile-mass range. The adopted mass prior should therefore be understood as an oscillation-motivated prior on the physical sterile-neutrino mass, rather than as a model-independent search for sterile states over arbitrary mass scales.
\subsection{Cosmological data}
\label{Cosmo_data}
We constrain the cosmological models considered in this work using a complementary combination of early- and late-time observations, including
the cosmic microwave background (CMB), baryon acoustic oscillations (BAO),
Type Ia supernovae (SNe Ia), redshift-space distortions (RSD), and the
Hubble Distance Network (H0DN) determination of the Hubble constant.
The individual data sets are described below.

\begin{enumerate}
\item \textbf{Cosmic Microwave Background (CMB).}
We use the Planck 2018 temperature and polarization likelihoods together with the reconstructed CMB lensing potential
\cite{PlanckLikelihood2018,Planck2018Lensing}. At low multipoles, we include the temperature and $E$-mode polarization likelihoods, \texttt{planck\_2018\_lowl.TT} and \texttt{planck\_2018\_lowl.EE}. At high multipoles, we adopt the
Planck 2018 CamSpec $TTTEEE$ likelihood, \texttt{planck\_2018\_highl\_CamSpec.TTTEEE}, including the $TT$, $TE$,
and $EE$ power spectra \cite{EfstathiouGratton2021}. We also include the Planck 2018 lensing-potential reconstruction through \texttt{planck\_2018\_lensing.clik} \cite{Planck2018Lensing}. The inclusion of CMB lensing is particularly relevant here, since massive neutrinos, spatial curvature, and dynamical dark energy all
affect the late-time growth of structure and the lensing potential, while BAO and SNe help break the associated geometric degeneracies. We refer to this complete combination as \textit{CMB} and use it consistently for all cosmological models and neutrino mass orderings.

\item \textbf{Baryon Acoustic Oscillations (DESI DR2).}
We use the DESI DR2 baryon acoustic oscillation measurements \cite{DESIDR2,DESI:2025zpo}, spanning the effective redshift range
$0.295\leq z\leq2.330$. The cosmological likelihood contains seven effective redshift bins and a total of 13 BAO distance measurements.
The lowest-redshift BGS bin provides the isotropic constraint $D_V/r_d$, while the remaining galaxy, quasar, and Ly$\alpha$ bins
provide anisotropic measurements of $D_M/r_d$ and $D_H/r_d$, where $D_M$ and $D_H=c/H(z)$ are the transverse comoving and Hubble
distances, respectively, and $r_d$ is the sound horizon at the baryon-drag epoch. The full covariance information supplied with the
DESI DR2 measurements is included in the likelihood. We denote this dataset by \textit{BAO}.

\item \textbf{Type Ia Supernovae (SNe Ia).}
Our fiducial supernova analysis uses the Pantheon+ compilation \cite{PantheonPlus}, containing 1701 light-curve measurements corresponding to 1550 distinct Type Ia supernovae extending to $z\simeq2.26$. We use the publicly available Pantheon+ likelihood implemented in \texttt{Cobaya}\footnote{\url{https://github.com/CobayaSampler/sn_data}}
\cite{Cobaya}, including the published statistical and systematic covariance information.

As a limited robustness check, we repeat the analysis of the CPL+$\Omega_k$+$(3+1)\nu$ model using Union3 \cite{Rubin:2023jdq} in place of Pantheon+, the resulting constraints are presented in Appendix~\ref{app:sn_robustness}.

\item \textbf{Redshift-space distortions (RSD).}
To complement the geometrical probes with information on the growth of large-scale structure, we use a conservative compilation of
redshift-space-distortion measurements of $f\sigma_8(z)$, spanning $0.067\leq z\leq1.52$. Nine measurements, listed in
Table~\ref{tab:fs8}, are treated as statistically independent. In addition, the three WiggleZ measurements at
$z=0.44$, $0.60$, and $0.73$ are included with their published covariance matrix \cite{Blake2012}.

For each model, the theoretical $f\sigma_8(z)$ values are obtained directly from \texttt{CLASS}, with the active and sterile neutrino
sectors included self-consistently in the linear perturbation evolution. We use the published marginalized $f\sigma_8$ measurements without applying an additional Alcock-Paczynski (AP) rescaling. Several measurements in the compilation were obtained from analyses that varied growth and AP parameters jointly \cite{Blake2012,GilMarin2017,Hou2018}. Applying an a posteriori AP correction to such compressed measurements would therefore be only approximate and would not reproduce the full covariance between growth and geometrical parameters. We consequently treat the published $f\sigma_8$ constraints as an effective compressed growth likelihood. Since the RSD contribution is subdominant to the CMB, BAO, and SNe Ia likelihoods in the present analysis, we retain the published measurements directly and regard any residual dependence on the survey fiducial
cosmology as a limitation of the compressed RSD treatment.

\begin{table}[t]
\caption{ Redshift-space-distortion measurements entering the $f\sigma_8$ likelihood and treated as statistically independent. The three correlated WiggleZ measurements included separately using their published covariance matrix.}
\label{tab:fs8}
\centering
\begin{tabular}{lccc}
\hline\hline
Survey & $z$ & $f\sigma_8(z)$ & Reference \\
\hline
6dFGS       & 0.067 & $0.4230\pm0.0550$ & \cite{Beutler2012} \\
2dFGRS      & 0.170 & $0.5100\pm0.0600$ & \cite{SongPercival2009} \\
BOSS LOWZ   & 0.320 & $0.3840\pm0.0950$ & \cite{ChuangLOWZ} \\
SDSS-II LRG & 0.370 & $0.4602\pm0.0378$ & \cite{Samushia2012} \\
BOSS CMASS  & 0.570 & $0.4260\pm0.0290$ & \cite{GilMarin2017} \\
VVDS        & 0.770 & $0.4900\pm0.1800$ & \cite{Guzzo2008,SongPercival2009} \\
VIPERS PDR2 & 0.860 & $0.4000\pm0.1100$ & \cite{Pezzotta2017} \\
FastSound   & 1.400 & $0.4820\pm0.1160$ & \cite{Okumura2016} \\
eBOSS DR14 QSO
            & 1.520 & $0.3960\pm0.0790$ & \cite{Hou2018} \\
\hline\hline
\end{tabular}
\end{table}

The WiggleZ measurements are collected into the data vector
\begin{equation}
\mathbf{d}_{\rm W}^{\rm obs}
= (0.413,0.390 ,0.437)^T, \, z=(0.44,\,0.60,\,0.73),
\end{equation}
with the covariance matrix
\begin{equation}
\mathbf{C}_{\rm W}
=
10^{-3}
\begin{pmatrix}
6.400 & 2.570 & 0     \\
2.570 & 3.969 & 2.540 \\
0     & 2.540 & 5.184
\end{pmatrix}.
\label{eq:wigglez_cov}
\end{equation}
The total growth likelihood is then written as
\begin{equation}
-2\ln{\cal L}_{f\sigma_8}
=
\chi^2_{\rm ind}+\chi^2_{\rm W},
\label{eq:rsd_likelihood}
\end{equation}
where
\begin{equation}
\chi^2_{\rm ind}=\sum_i\frac{\left[f\sigma_{8,i}^{\rm obs}
-
f\sigma_{8,i}^{\rm th}\right]^2}{\sigma_{f\sigma_8,i}^{\,2}},
\end{equation}
and $\chi^2_{\rm W}=\Delta\mathbf{d}_{\rm W}^{\,T}
\mathbf{C}_{\rm W}^{-1}
\Delta\mathbf{d}_{\rm W},
$ with $\Delta\mathbf{d}_{\rm W}=\mathbf{d}_{\rm W}^{\rm obs}-\mathbf{d}_{\rm W}^{\rm th}.$

\item \textbf{Hubble Distance Network (H0DN).}

We include the recent Hubble Constant Distance Network (H0DN) determination as an external constraint on the
late-time absolute expansion scale \cite{H0DN2026}. The H0DN analysis combines multiple independent distance indicators and finds an internally consistent determination that is robust against the removal of individual calibrators. We adopt
\begin{equation}
H_0^{\rm H0DN}
=
73.50\pm0.81~
{\rm km\,s^{-1}\,Mpc^{-1}},
\label{eq:H0DN}
\end{equation}
and implement it through the Gaussian likelihood
\begin{equation}
-2\ln{\cal L}_{H_0}
=
\frac{\left(H_0-73.50\right)^2}{0.81^2}.
\label{eq:H0DN_like}
\end{equation}

The H0DN constraint is particularly relevant for the extended cosmologies considered here, since the inferred value of $H_0$ can be correlated with the sterile-neutrino abundance and mass sector, the dynamical dark-energy parameters, and spatial curvature. It therefore provides complementary absolute-scale 
\end{enumerate}

For the fiducial analysis, we adopt the combined data set
\begin{equation}
\mathcal{D}_{\rm PP}={\rm CMB}+{\rm BAO}_{\rm DESI\,DR2}+{\rm PP}
+{\rm RSD}+H_0^{\rm H0DN},
\label{eq:data_PP}
\end{equation}
where PP denotes the Pantheon+ SNe Ia compilation. Unless stated
otherwise, all results presented in the main text are obtained using
$\mathcal{D}_{\rm PP}$. 
\section{Statistical Analysis}
\label{sec:stat}
\subsection{Model space and parameter priors}
\label{subsec:models_priors}
We consider four cosmological model classes, summarized in Table~\ref{tab:models}, each analyzed separately for normal and inverted ordering of the three active-neutrino masses. All models share the same $(3+1)\nu$ sector, while the late-time background is varied through the dark-energy and spatial-curvature sectors.

The flat $\Lambda$CDM$+(3+1)\nu$ model serves as the reference cosmology, with $w=-1$ and $\Omega_k=0$. We then consider the flat CPL$+(3+1)\nu$ extension, in which the dark-energy equation of state is allowed to evolve, and the $\Lambda$CDM$+\Omega_k+(3+1)\nu$ model, which retains $w=-1$ while allowing spatial curvature to vary. Finally,
we consider the CPL$+\Omega_k+(3+1)\nu$ model, in which both dynamical dark energy and spatial curvature are free. The
$\Lambda$CDM$+\Omega_k+(3+1)\nu$ case provides a useful control model for isolating the effect of curvature from that of CPL dynamics.

\begin{table}[t]
\caption{Cosmological model classes considered in this work. Each model is analysed independently for normal and inverted ordering of the active-neutrino masses.}
\label{tab:models}
\centering
\footnotesize

\begin{tabular}{llll}
\hline\hline
Model
& Geometry
& Dark energy
& Neutrino sector
\\
\hline

$\Lambda{\rm CDM}+(3+1)\nu$
& flat
& $w=-1$
& $(3+1)\nu$, NO/IO
\\

${\rm CPL}+(3+1)\nu$
& flat
& CPL
& $(3+1)\nu$, NO/IO
\\

$\Lambda{\rm CDM}+\Omega_k+(3+1)\nu$
& non-flat
& $w=-1$
& $(3+1)\nu$, NO/IO
\\

${\rm CPL}+\Omega_k+(3+1)\nu$
& non-flat
& CPL
& $(3+1)\nu$, NO/IO
\\

\hline\hline
\end{tabular}
\end{table}

The principal sampling priors are listed in Table~\ref{tab:priors}. All priors are uniform within the quoted ranges, and identical priors are adopted for NO and IO. The same prior choices are retained for the different observational data combinations considered in Sec.~\ref{sec:data}. This allows the shifts in the posterior distributions to be attributed directly to changes in the data rather than to changes in prior volume.

\begin{table}[t]
\caption{
Uniform sampling priors adopted in the cosmological analysis.
The curvature parameter $\Omega_k$ is varied only in the non-flat
models.
}
\label{tab:priors}
\centering
\begin{tabular}{lr}
\hline\hline
Parameter & Prior \\
\hline
$m_{\rm lightest}$ [eV] & $\mathcal U[0.001,\,0.20]$ \\
$m_s$ [eV]              & $\mathcal U[0.50,\,1.50]$ \\
$\Delta N_{\rm eff}$    & $\mathcal U[0.01,\,0.80]$ \\
$w_0$                   & $\mathcal U[-2.0,\,-0.4]$ \\
$w_a$                   & $\mathcal U[-2.0,\,1.0]$ \\
$\Omega_k$              & $\mathcal U[-0.05,\,0.05]$ \\
$H_0$ [${\rm km/s/Mpc}$] & $\mathcal U[60,\,80]$ \\
$\omega_b$              & $\mathcal U[0.019,\,0.025]$ \\
$\omega_{\rm c}$      & $\mathcal U[0.09,\,0.16]$ \\
$\tau_{\rm reio}$       & $\mathcal U[0.01,\,0.15]$ \\
$n_s$                    & $\mathcal U[0.90,\,1.05]$ \\
$\ln(10^{10}A_s)$       & $\mathcal U[1.61,\,3.91]$ \\
\hline\hline
\end{tabular}
\end{table}

The sterile-neutrino mass prior, $0.5\leq m_s/{\rm eV}\leq1.5$, is chosen to target the eV-scale sterile-neutrino region motivated by short-baseline oscillation phenomenology, as discussed in Sec \ref{Nu_Osc_data}. The sterile abundance is varied independently through $\Delta N_{\rm eff}$ (see Sec~\ref{subsec_str_nu}). We restrict the analysis to $0.01\leq\Delta N_{\rm eff}\leq0.80$. This range restricts the analysis to partially thermalized sterile-neutrino scenarios, excluding both the exactly unpopulated limit, $\Delta N_{\rm eff}=0$, and the fully thermalized case, $\Delta N_{\rm eff}=1$. Our analysis should therefore be interpreted as a cosmological test of an oscillation-motivated partially thermalized eV-scale sterile state, rather than as a model-independent search over arbitrary sterile-neutrino masses.

Since $m_{s,\rm eff}$ is a derived parameter, the uniform priors on $m_s$ and $\Delta N_{\rm eff}$ induce a non-uniform prior on $m_{s,\rm eff}$. We obtain this prior from 
\begin{multline} \label{eq:induced_prior}
\pi(m_{s,\rm eff})=\int dm_s\,d(\Delta N_{\rm eff})\,\pi(m_s,\Delta N_{\rm eff})
\\ \times\delta_{\rm D}\!\left[m_{s,\rm eff}-m_s(\Delta N_{\rm eff})^{3/4}\right],
\end{multline}
and normalize it to unit integral before comparing it with the marginalized posterior distributions discussed in Sec.~\ref{subsec:interplay}.

\subsection{Numerical implementation and likelihood}
\label{subsec:numerical}

The background evolution and linear perturbations are calculated using \texttt{CLASS}. A custom \texttt{CLASS-Cobaya} interface maps $\left(m_{\rm lightest},\,m_s,\,\Delta N_{\rm eff},\,{\rm ordering}\right) $ onto the four physical non-cold relic masses and temperatures before each Boltzmann evaluation.

Posterior sampling is performed with \texttt{Cobaya} using multiple parallel Markov chains and adaptive proposal learning. Each cosmological model and active-neutrino mass ordering is sampled independently. Convergence is monitored using the Gelman-Rubin statistic, and the final chains are required to satisfy $R-1 \lesssim 0.03$. For the fiducial Pantheon+ data combination $\mathcal{D}_{\rm PP}$ defined in Eq.~\eqref{eq:data_PP}, the total log-likelihood is
\begin{align}
\ln{\cal L}_{\rm tot}
={}&
\ln{\cal L}_{\rm CMB}
+\ln{\cal L}_{\rm BAO}
+\ln{\cal L}_{\rm PP}
\nonumber\\
&+
\ln{\cal L}_{f\sigma_8}
+\ln{\cal L}_{H_0}.
\label{eq:total_likelihood}
\end{align}

Marginalized posterior distributions and credible intervals are obtained from the converged chains using \texttt{GetDist} \cite{LewisGetDist}. The quoted two-sided uncertainties correspond to 68\% marginalized credible intervals, while one-sided upper bounds are reported at 95\% confidence.

\subsection{Goodness of fit and model comparison}
\label{subsec:modelcomparison_method}
We compare each extended model with the corresponding flat $\Lambda$CDM+$(3+1)\nu$ reference model having the same active-neutrino mass ordering. Thus, NO and IO models are assessed independently, allowing the effects of dark-energy and curvature freedom to be separated from those associated with the mass ordering.

The improvement in the best-fit likelihood is quantified by
\begin{equation}
\Delta\chi^2_{\rm min}\equiv\chi^2_{{\rm min},i}-\chi^2_{{\rm min},{\rm ref}},
\label{eq:deltachi2}
\end{equation}
where $\chi^2_{\rm min}=-2\ln{\cal L}_{\rm max}$ and 'ref' denotes the ordering-matched flat $\Lambda$CDM+$(3+1)\nu$ model.  With this convention, $\Delta\chi^2_{\rm min}<0$ indicates that the extended model provides a better maximum-likelihood fit than its corresponding reference cosmology.

To account for model complexity, we use the Akaike and Bayesian information criteria,
\begin{equation}
{\rm AIC}=\chi^2_{\rm min}+2k,
\qquad
{\rm BIC}=\chi^2_{\rm min}+k\ln N,
\label{eq:IC}
\end{equation}
where $k$ is the number of fitted parameters and $N$ is the total number of data points. Their differences are defined as
\begin{equation}
\Delta{\rm IC} = {\rm IC}_i-{\rm IC}_{\rm ref}, \qquad
{\rm IC}\in\{{\rm AIC},{\rm BIC}\}.
\label{eq:deltaIC}
\end{equation}
Negative values favor the extended model, while positive values favor the ordering-matched reference model.  

As a complementary model-selection statistic, we compute the Bayesian evidence using \texttt{MCEvidence} \cite{Heavens:2017afc,MCEvidenceGitHub}. For a model ${\cal M}$ with parameters $\boldsymbol{\theta}$, the Bayesian evidence is
\begin{equation}
{\cal Z}_{\cal M}
\equiv(D|{\cal M})=\int{\cal L}(D|\boldsymbol{\theta},{\cal M})\, \pi(\boldsymbol{\theta}|{\cal M})\,
d\boldsymbol{\theta},
\label{eq:evidence}
\end{equation}
where ${\cal L}$ is the likelihood and $\pi$ denotes the prior density. We report the logarithmic evidence, $\ln{\cal Z}$, and the Bayes factor of an extended model ${\cal M}_i$ relative to the ordering-matched reference model
\begin{equation}
B_{i,\rm ref}=\frac{{\cal Z}_i}{{\cal Z}_{\rm ref}},
\end{equation}

or equivalently

\begin{equation}
\ln B_{i,\rm ref}=\ln{\cal Z}_i-\ln{\cal Z}_{\rm ref} .
\label{eq:bayesfactor}
\end{equation}
With this convention, $\ln B_{i,\rm ref}>0$ favors the extended model,
whereas $\ln B_{i,\rm ref}<0$ favors the ordering-matched reference
model.

These statistics probe complementary aspects of model performance. While $\Delta\chi^2_{\rm min}$ measures the best-fit improvement, AIC and BIC penalize additional parameters, and the Bayesian evidence averages over the full prior volume. Hence, improved fit quality does not necessarily imply comparable Bayesian support.

Since the flat CPL+$(3+1)\nu$ model is nested within its non-flat extension at $\Omega_k=0$, we additionally use the Savage-Dickley density ratio as a nested-model cross-check of the Bayesian evidence for spatial curvature \cite{VerdinelliWasserman1995,Trotta2007}. Provided that the priors for all parameters common to the two models are identical, the Bayes factor for the flat model relative to the non-flat model is
\begin{equation}
B_{\rm flat,nonflat}^{\rm SD}=\frac{p(\Omega_k=0\mid D,\mathcal{M}_{\rm nonflat})}
{\pi(\Omega_k=0\mid\mathcal{M}_{\rm nonflat})}.
\label{eq:SD_curvature}
\end{equation}
Thus, $B_{\rm flat,nonflat}^{\rm SD}>1$ favors the flat reference model,
whereas $B_{\rm flat,nonflat}^{\rm SD}<1$ favors the non-flat extension. We evaluate this quantity separately for NO and IO, always comparing the flat and non-flat realizations of the same CPL$+(3+1)\nu$ framework.

An analogous nested comparison is performed for the dark-energy sector.  The non-flat $\Lambda+(3+1)\nu$ model is recovered from the non-flat CPL+$(3+1)\nu$ model at $(w_0,w_a)=(-1,0)$.  In this case the relevant Savage-Dickery ratio involves the joint marginalized density,
\begin{equation}
B_{\Lambda,{\rm CPL}}^{\rm SD} = \frac{p(w_0=-1,w_a=0\mid D,\mathcal{M}_{\rm CPL})}
{\pi(w_0=-1,w_a=0\mid\mathcal{M}_{\rm CPL})}.
\label{eq:SD_darkenergy}
\end{equation}

We do not use the Savage-Dickey ratio to compare the full extended model directly with $\Lambda$CDM.  In particular, the no-sterile limit is not a regular nested point of the adopted parameterization, since the prior imposes $\Delta N_{\rm eff}>0$ and the sterile-mass parameter becomes unidentified as the sterile abundance vanishes.

\section{Results and Discussion}
\label{sec:results}

We now present the posterior constraints for the four model classes defined in Table~\ref{tab:models}, using the fiducial data combination $\mathcal{D}_{\rm PP}$.  Each model is analyzed independently for NO and IO.  Marginalized uncertainties correspond to $68\%$ credible intervals, while one-sided upper limits are quoted at $95\%$ credibility.  We first examine the progression from the flat $\Lambda$CDM+$(3+1)\nu$ reference model to the CPL and CPL+$\Omega_k$ extensions, and subsequently use the $\Lambda$CDM+$\Omega_k$ model to disentangle the effects of spatial
curvature from those of dynamical dark energy.

\subsection{Cosmological parameter constraints}
\label{subsec:cosmo_results}

\begin{figure*}[t]
    \centering

    \begin{minipage}[t]{0.49\textwidth}
        \centering
        \includegraphics[width=\linewidth]{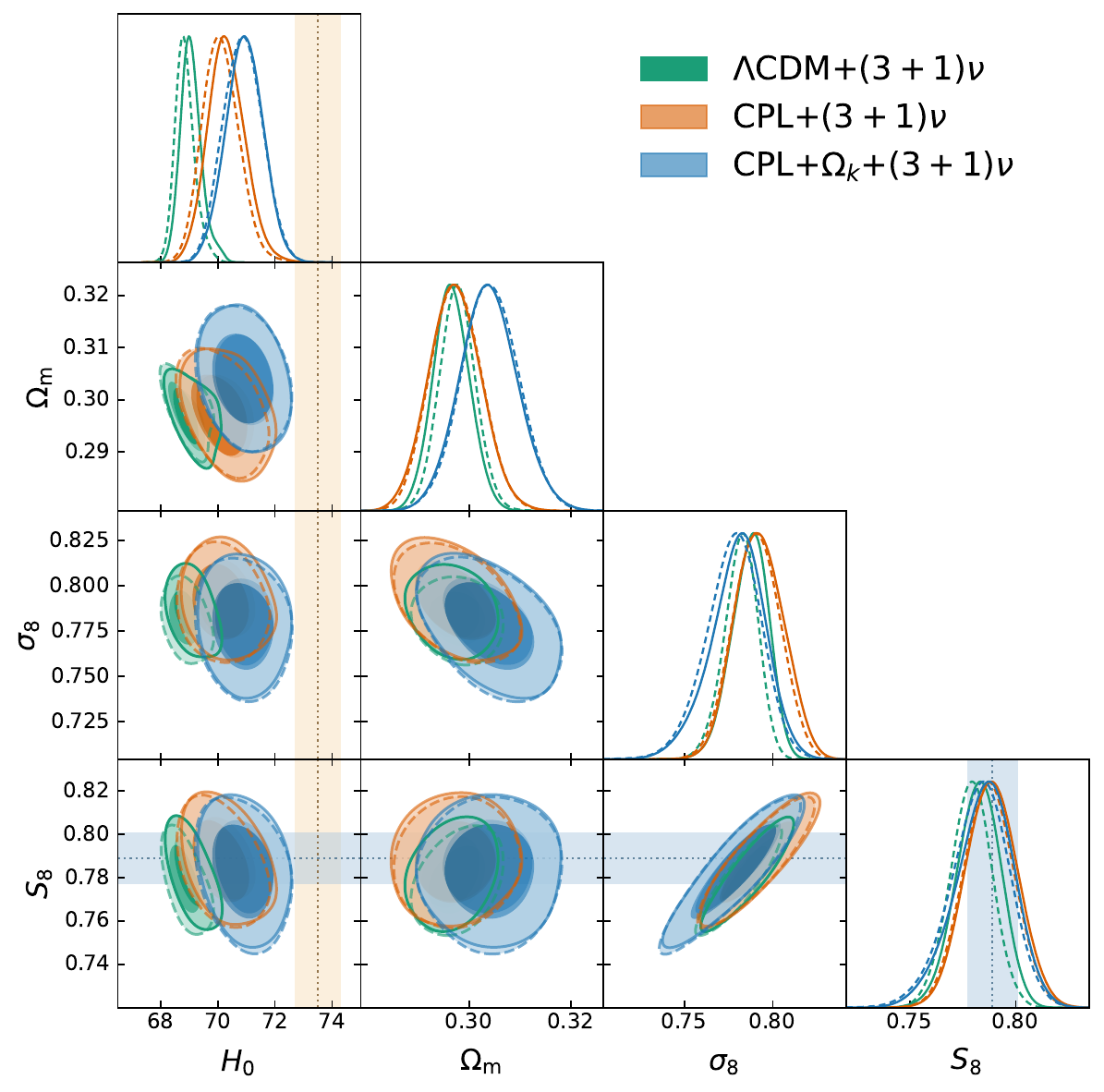}
        \vspace{-2mm}
        {\small\textbf{(a)}}
    \end{minipage}
    \hfill
    \begin{minipage}[t]{0.48\textwidth}
        \centering
        \includegraphics[width=\linewidth]{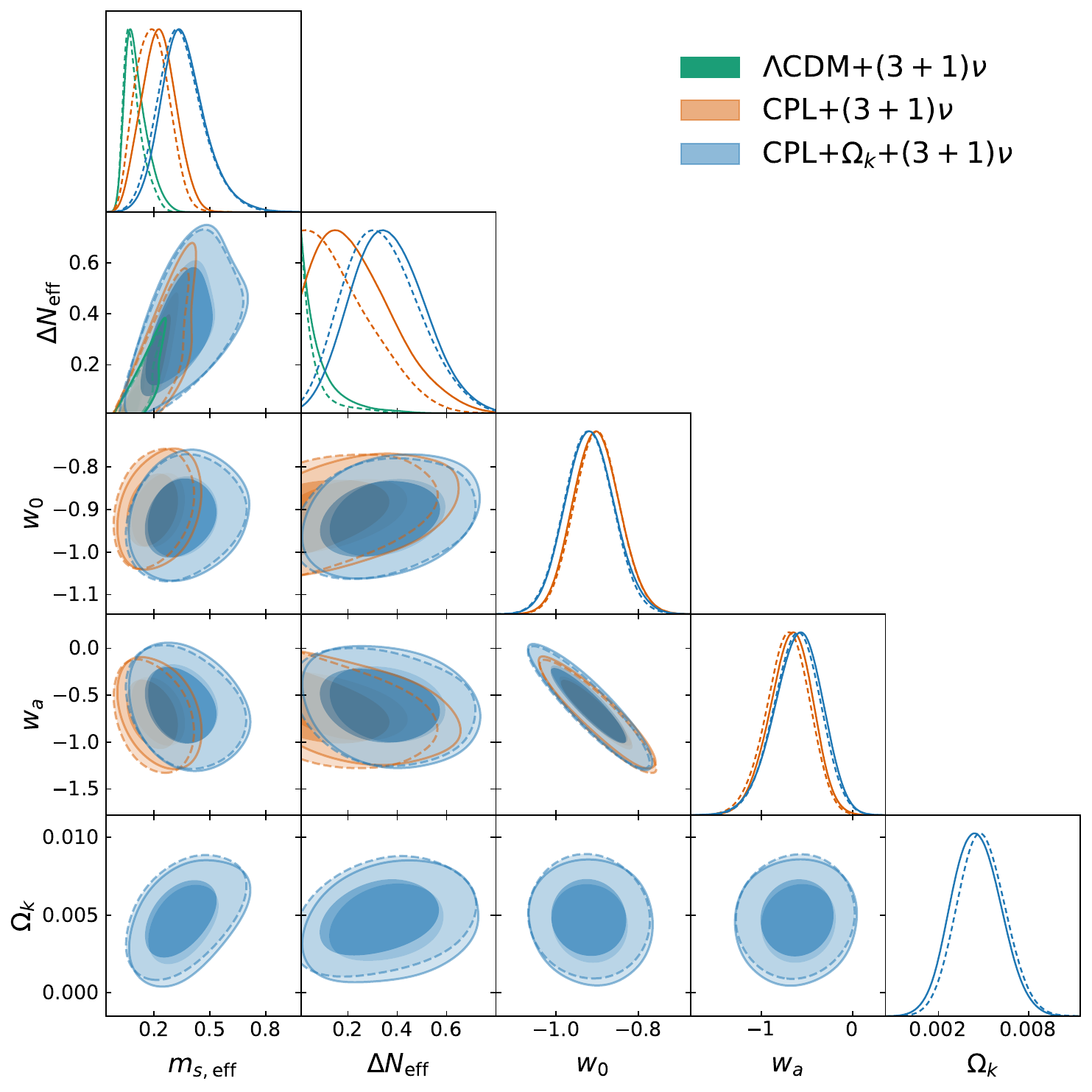}
        \vspace{-2mm}
        {\small\textbf{(b)}}
    \end{minipage}

    \caption{Marginalized one- and two-dimensional posterior distributions for the cosmological models considered in this work. \textbf{(a)} Constraints in
the $(H_0,\Omega_{\rm m},\sigma_8,S_8)$ subspace for the flat $\Lambda$CDM+$ (3+1)\nu$, flat CPL+$ (3+1)\nu$, and CPL+$\Omega_k$+$ (3+1)\nu$ models, each analyzed for normal and inverted mass ordering. The shaded vertical band in the $H_0$ panels shows the $68\%$ H0DN constraint, $H_0=73.50\pm0.81~{\rm km\,s^{-1}\,Mpc^{-1}}$, which is included in the fiducial likelihood and is displayed here only as a visual reference to the adopted late-time expansion constraint. The shaded horizontal band
in the $S_8$ panels denotes the $68\%$ DES Y6 constraint, $S_8=0.789\pm0.012$ \cite{DESY6}, this measurement is not included in the likelihood and is shown only as an external reference. \textbf{(b)} Joint constraints in the $(m_{s,\rm eff},\Delta N_{\rm eff},w_0,w_a,\Omega_k)$ subspace, illustrating correlations among the sterile-neutrino sector, dynamical dark energy, and spatial curvature. Panels involving $\Omega_k$ apply only to the non-flat CPL+$\Omega_k$ extensions. Inner and outer contours denote the $68\%$ and $95\%$ credible regions, respectively.}
\label{fig:combined_posteriors}
\end{figure*}

The marginalized limits are summarized in Table~\ref{tab:constraints} and principal posterior constraints are shown in Fig.~\ref{fig:combined_posteriors}. Panel (a) displays the late-time parameters $H_0$, $\Omega_{\rm m}$, $\sigma_8$, and $S_8$, while panel (b) focuses on the sterile-neutrino, dark-energy, and curvature sectors.

\begin{table*}[t]
\caption{Marginalized constraints from the fiducial data combination $\mathcal{D}_{\rm PP}={\rm CMB}+{\rm BAO}+{\rm PP}+{\rm RSD}+{\rm H0DN}$. Quoted uncertainties correspond to $68\%$ credible intervals, while one-sided upper limits are given at $95\%$ credibility.
$H_0$ is in ${\rm km\,s^{-1}\,Mpc^{-1}}$ and neutrino masses are in eV. The limits on the physical sterile-neutrino mass $m_s$ are conditional on the adopted prior $0.5\leq m_4/{\rm eV}\leq1.5$. For the $\Lambda$CDM models, $(w_0,w_a)=(-1,0)$ are fixed,
while $\Omega_k=0$ is fixed in all flat models.}
\label{tab:constraints}

\footnotesize
\centering
\setlength{\tabcolsep}{3.2pt}
\renewcommand{\arraystretch}{1.02}

\begin{ruledtabular}
\begin{tabular}{lllllll}
& \multicolumn{2}{c}{$\Lambda$CDM+$ (3+1)\nu$}
& \multicolumn{2}{c}{CPL+$ (3+1)\nu$}
& \multicolumn{2}{c}{CPL+$\Omega_k$+$ (3+1)\nu$}
\\
Parameter
& NO & IO
& NO & IO
& NO & IO
\\
\hline

$H_0$
& $69.06^{+0.31}_{-0.40}$
& $68.85^{+0.27}_{-0.38}$
& $70.30^{+0.60}_{-0.68}$
& $70.10^{+0.58}_{-0.67}$
& $70.93\pm0.65$
& $70.88\pm0.67$
\\

$\omega_b$
& $0.02268\pm0.00015$
& $0.02270\pm0.00015$
& $0.02259^{+0.00015}_{-0.00018}$
& $0.02256\pm0.00017$
& $0.02244\pm0.00017$
& $0.02241\pm0.00017$
\\

$\omega_c$
& $0.11685^{+0.00066}_{-0.0012}$
& $0.11628^{+0.00062}_{-0.0011}$
& $0.1211^{+0.0017}_{-0.0028}$
& $0.1203^{+0.0014}_{-0.0025}$
& $0.1256^{+0.0025}_{-0.0029}$
& $0.1253^{+0.0025}_{-0.0031}$
\\

$\tau_{\rm reio}$
& $0.0660^{+0.0079}_{-0.0093}$
& $0.0687^{+0.0080}_{-0.0095}$
& $0.0593^{+0.0073}_{-0.0085}$
& $0.0601^{+0.0072}_{-0.0088}$
& $0.0585^{+0.0070}_{-0.0086}$
& $0.0589^{+0.0072}_{-0.0088}$
\\

$n_s$
& $0.9762^{+0.0038}_{-0.0049}$
& $0.9767^{+0.0036}_{-0.0047}$
& $0.9749^{+0.0055}_{-0.0069}$
& $0.9739^{+0.0047}_{-0.0068}$
& $0.9706^{+0.0060}_{-0.0070}$
& $0.9694^{+0.0060}_{-0.0069}$
\\

$\ln(10^{10}A_s)$
& $3.062^{+0.016}_{-0.020}$
& $3.067^{+0.016}_{-0.019}$
& $3.057^{+0.016}_{-0.019}$
& $3.057^{+0.015}_{-0.019}$
& $3.066^{+0.016}_{-0.018}$
& $3.066^{+0.016}_{-0.019}$
\\

\hline

$m_{\rm lightest}$
& $<0.00880$
& $<0.0101$
& $<0.0133$
& $<0.0156$
& $<0.0232$
& $<0.0253$
\\

$m_s$
& $<0.891$
& $<0.932$
& $<0.782$
& $<0.807$
& $<0.889$
& $<0.924$
\\

$\Delta N_{\rm eff}$
& $<0.0956$
& $<0.0672$
& $0.243^{+0.075}_{-0.210}$
& $<0.244$
& $0.36^{+0.14}_{-0.16}$
& $0.34^{+0.13}_{-0.17}$
\\

$w_0$
& $-1$ (fixed)
& $-1$ (fixed)
& $-0.902\pm0.056$
& $-0.900\pm0.055$
& $-0.917\pm0.059$
& $-0.921\pm0.058$
\\

$w_a$
& $0$ (fixed)
& $0$ (fixed)
& $-0.67^{+0.25}_{-0.22}$
& $-0.71^{+0.25}_{-0.22}$
& $-0.60^{+0.28}_{-0.24}$
& $-0.61^{+0.27}_{-0.24}$
\\

$\Omega_k$
& $0$ (fixed)
& $0$ (fixed)
& $0$ (fixed)
& $0$ (fixed)
& $0.0045\pm0.0016$
& $0.0049\pm0.0016$
\\

\hline

$\sum m_{\nu,\rm active}$
& $0.0707^{+0.0016}_{-0.011}$
& $0.1097^{+0.0017}_{-0.0097}$
& $0.0780^{+0.0036}_{-0.0180}$
& $0.1169^{+0.0040}_{-0.0170}$
& $0.0982^{+0.0059}_{-0.0390}$
& $0.1315^{+0.0064}_{-0.0320}$
\\

$m_{s,\rm eff}$
& $0.106^{+0.030}_{-0.075}$
& $0.090^{+0.022}_{-0.064}$
& $0.226\pm0.090$
& $0.192^{+0.080}_{-0.098}$
& $0.361^{+0.095}_{-0.130}$
& $0.350^{+0.100}_{-0.140}$
\\

$\Omega_{\rm m}$
& $0.2965\pm0.0036$
& $0.2977\pm0.0036$
& $0.2972\pm0.0051$
& $0.2974\pm0.0049$
& $0.3039\pm0.0056$
& $0.3041\pm0.0055$
\\

$\sigma_8$
& $0.787^{+0.011}_{-0.0093}$
& $0.782^{+0.010}_{-0.0084}$
& $0.792^{+0.013}_{-0.015}$
& $0.791\pm0.013$
& $0.780^{+0.017}_{-0.013}$
& $0.778^{+0.017}_{-0.014}$
\\

$S_8$
& $0.782\pm0.010$
& $0.7789\pm0.0099$
& $0.789\pm0.012$
& $0.787\pm0.012$
& $0.785^{+0.015}_{-0.012}$
& $0.783^{+0.015}_{-0.012}$
\\

\hline

$\chi^2_{\rm min}$
& $13349.7$
& $13355.0$
& $13336.12$
& $13339.20$
& $13328.59$
& $13330.24$
\\

\end{tabular}
\end{ruledtabular}
\end{table*}

For the flat $\Lambda$CDM+$(3+1)\nu$ reference model we obtain $H_0 = 69.06^{+0.31}_{-0.40} \ {\rm km\,s^{-1}\,Mpc^{-1}} \, ({\rm NO}), 
H_0 = 68.85^{+0.27}_{-0.38} \ {\rm km\,s^{-1}\,Mpc^{-1}}\, ({\rm IO})$ . Allowing a dynamical CPL equation of state shifts these values upward to $70.30^{+0.60}_{-0.68}$ and $70.10^{+0.58}_{-0.67}\ {\rm km\,s^{-1}\,Mpc^{-1}}$ for NO and IO, respectively.  When spatial curvature is additionally varied, the posterior moves further to $70.93\pm0.65$ and $70.88\pm0.67\ {\rm km\,s^{-1}\,Mpc^{-1}}$.  Thus, the dominant shift in the inferred expansion rate is associated with the additional background freedom rather than with the choice of neutrino mass ordering.

The matter density is comparatively insensitive to the introduction of CPL dynamics, both the
$\Lambda$CDM+$(3+1)\nu$ and CPL+$(3+1)\nu$ models give $\Omega_{\rm m}\simeq0.297$.  Allowing curvature instead shifts the
CPL+$\Omega_k$ models to $\Omega_{\rm m}\simeq0.304$. The clustering amplitude remains similarly stable.  We find
$S_8=0.782\pm0.010$ and $0.7789\pm0.0099$ for the $\Lambda$CDM+$(3+1)\nu$ NO and IO models,
$0.789\pm0.012$ and $0.787\pm0.012$ for the corresponding CPL models, and $0.785^{+0.015}_{-0.012}$ and
$0.783^{+0.015}_{-0.012}$ when curvature is also included. The strong overlap of the NO and IO posteriors across these model classes shows that the late-time cosmological constraints depend much more strongly on the assumed background dynamics and geometry than on the active-neutrino mass ordering.

The same hierarchy is reflected in the sterile sector.  The effective sterile contribution increases from
$m_{s,\rm eff}=0.106^{+0.030}_{-0.075}\ {\rm eV}$ (NO) and $0.090^{+0.022}_{-0.064}\ {\rm eV}$ (IO) in the
$\Lambda$CDM+$(3+1)\nu$ reference model to $0.226\pm0.090$ and $0.192^{+0.080}_{-0.098}\ {\rm eV}$ in the CPL extension, and to $0.361^{+0.095}_{-0.130}$ and $0.350^{+0.100}_{-0.140}\ {\rm eV}$ in the CPL+$\Omega_k$ model.  A corresponding broadening toward larger $\Delta N_{\rm eff}$ accompanies these shifts.  This progression already indicates that the inferred sterile-neutrino properties are correlated with the freedom allowed in the late-time background,
an interplay examined in more detail in the following subsection.
\subsection{Interplay of spatial curvature, dark energy, and the sterile sector}
\label{subsec:interplay}

\begin{figure}[h!]
\centering
\includegraphics[width=0.9\columnwidth]{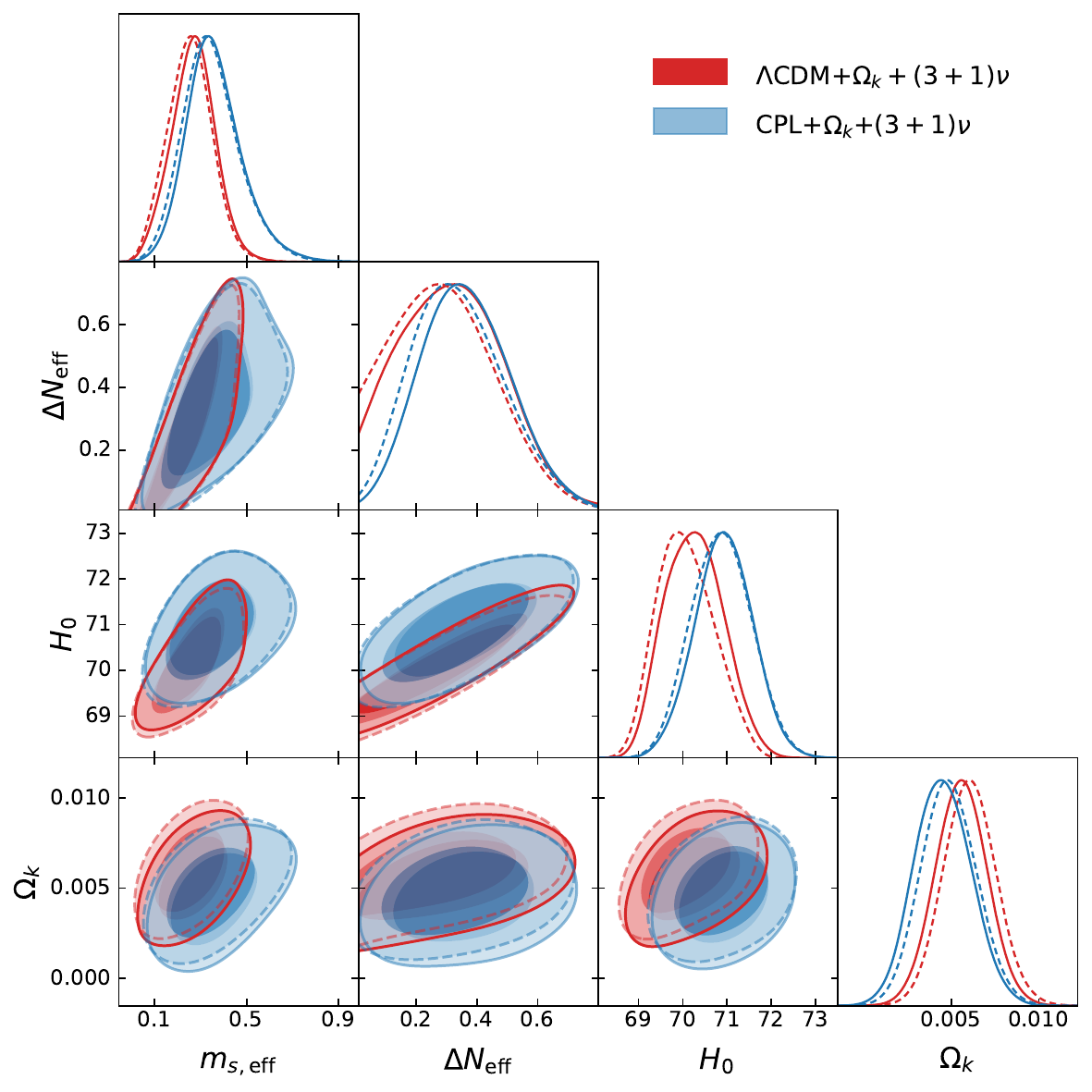}
\caption{
Marginalized posterior distributions for the non-flat CPL+$ (3+1)\nu$ and non-flat $\Lambda+(3+1)\nu$ models under
normal and inverted ordering, using $\mathcal{D}_{\rm PP}$.  The comparison isolates the effect of fixing the dark-energy sector to
$(w_0,w_a)=(-1,0)$ while retaining spatial curvature and the same neutrino-sector assumptions.
}
\label{fig:dde_contours}
\end{figure}

The model hierarchy introduced above allows the effects of spatial curvature and dynamical dark energy to be separated.  In particular, the $\Lambda$CDM+$\Omega_k$+$(3+1)\nu$ model provides a useful control for the CPL+$\Omega_k$+$(3+1)\nu$ model, the two share the same curvature and neutrino sectors, while the former is obtained by fixing $(w_0,w_a)=(-1,0)$.  The marginalized constraints for the $\Lambda$CDM+$\Omega_k$+$(3+1)\nu$ models are summarized in
Table~\ref{tab:nonflat_lambda_constraints}, and the corresponding joint posterior comparison is shown in Fig.~\ref{fig:dde_contours}.

\begin{table}[t]
\caption{Marginalized constraints for the $\Lambda+\Omega_k+(3+1)\nu$ models for NO and IO using $\mathcal{D}_{\rm PP}$. The dark-energy parameters are fixed to
$(w_0,w_a)=(-1,0)$.}
\label{tab:nonflat_lambda_constraints}
\footnotesize
\centering

\begin{ruledtabular}
\begin{tabular}{lll}
Parameter & NO & IO \\
\hline
$H_0$ 
& $70.26^{+0.62}_{-0.71}$ 
& $70.08^{+0.60}_{-0.72}$ \\

$\omega_b$
& $0.02251\pm 0.00017$
& $0.02249\pm0.00017$ \\

$\omega_c$
& $0.1241\pm 0.0029 $ 
& $0.1239\pm0.0029$ \\

$\tau_{\rm reio}$
& $0.0634^{+0.0079}_{-0.0093}$
& $0.0644^{+0.0079}_{-0.0094}$ \\

$n_s$
& $0.9727\pm 0.0064$ 
& $0.9718^{+0.0059}_{-0.0066}$ \\

$\ln(10^{10}A_s)$
& $3.073^{+0.018}_{-0.021} $ 
& $3.075^{+0.018}_{-0.021}$ \\
\hline
$m_{\rm lightest}$ 
& $< 0.0137 $
& $<0.0149$ \\

$m_s$ 
& $< 0.707 $ 
& $<0.713$ \\

$\Delta N_{\rm eff}$
& $0.32^{+0.16}_{-0.19} $
& $0.30^{+0.14}_{-0.20}$ \\

$\Omega_k$
& $0.0056\pm 0.0015 $
& $0.0060\pm0.0015$ \\
\hline
$\sum m_{\nu,\rm active}$ 
& $0.0789^{+0.0035}_{-0.019}$ 
& $0.1162^{+0.0026}_{-0.016}$ \\
$m_{s,\rm eff}$
& $0.269\pm 0.096 $
& $0.255\pm0.098$ \\
$\Omega_m$
& $0.3045\pm 0.0043 $ 
& $0.3061\pm0.0045$ \\

$\sigma_8$
& $0.778\pm 0.012  $ 
& $0.774\pm0.013$ \\

$S_8$
&  $0.784^{+0.012}_{-0.011}$ 
& $0.781^{+0.013}_{-0.011}$ \\
\hline
$\chi^2_{\rm min}$
& $13336.0$
& $13338.8$

\end{tabular}
\end{ruledtabular}

\end{table}
At fixed curved geometry, allowing CPL dynamics produces a correlated shift in the expansion rate and spatial curvature.  For NO, the $\Lambda$CDM+$\Omega_k$+$(3+1)\nu$ model gives
$
H_0=70.26^{+0.62}_{-0.71}\ {\rm km\,s^{-1}\,Mpc^{-1}},
\qquad
\Omega_k=0.0056\pm0.0015 ,
$
whereas the CPL+$\Omega_k$+$(3+1)\nu$ model gives
$
H_0=70.93\pm0.65\ {\rm km\,s^{-1}\,Mpc^{-1}},
\qquad
\Omega_k=0.0045\pm0.0016 .
$
For IO, the corresponding change is from
$
H_0=70.08^{+0.60}_{-0.72}\ {\rm km\,s^{-1}\,Mpc^{-1}},
\qquad
\Omega_k=0.0060\pm0.0015 ,
$
to
$
H_0=70.88\pm0.67\ {\rm km\,s^{-1}\,Mpc^{-1}},
\qquad
\Omega_k=0.0049\pm0.0016 .
$
Thus, once curvature is allowed, introducing CPL freedom shifts the posterior toward slightly larger $H_0$ and smaller positive $\Omega_k$.  The similar behaviour for NO and IO indicates that this shift is primarily associated with the late-time background
parameterization rather than with the active-neutrino mass ordering.

The sterile sector changes coherently with the background geometry.
For the $\Lambda$CDM+$\Omega_k$ models we obtain
$
\Delta N_{\rm eff}
 =0.32^{+0.16}_{-0.19},\qquad
m_{s,\rm eff}=0.269\pm0.096~{\rm eV}
$
for NO, and
$
\Delta N_{\rm eff}
 =0.30^{+0.14}_{-0.20},\qquad
m_{s,\rm eff}=0.255\pm0.098~{\rm eV}
$
for IO.  Allowing CPL dynamics shifts these constraints to
$
\Delta N_{\rm eff}
 =0.36^{+0.14}_{-0.16},\qquad
m_{s,\rm eff}
 =0.361^{+0.095}_{-0.130}~{\rm eV}
$
for NO and
$
\Delta N_{\rm eff}
 =0.34^{+0.13}_{-0.17},\qquad
m_{s,\rm eff}
 =0.350^{+0.100}_{-0.140}~{\rm eV}$
for IO.  The CPL+$\Omega_k$ models therefore accommodate a somewhat larger effective sterile contribution than their $\Lambda$CDM+$\Omega_k$ counterparts.  Together with the curvature shift above, this indicates a joint degeneracy among the expansion history, spatial geometry, and the cosmological sterile-neutrino abundance. These correlations are visible directly in Fig.~\ref{fig:dde_contours}, where the $\Lambda$CDM+$\Omega_k$ and CPL+$\Omega_k$ posteriors are compared for both mass orderings.  

\begin{figure*}[t]
\centering
\includegraphics[width=0.9\textwidth]{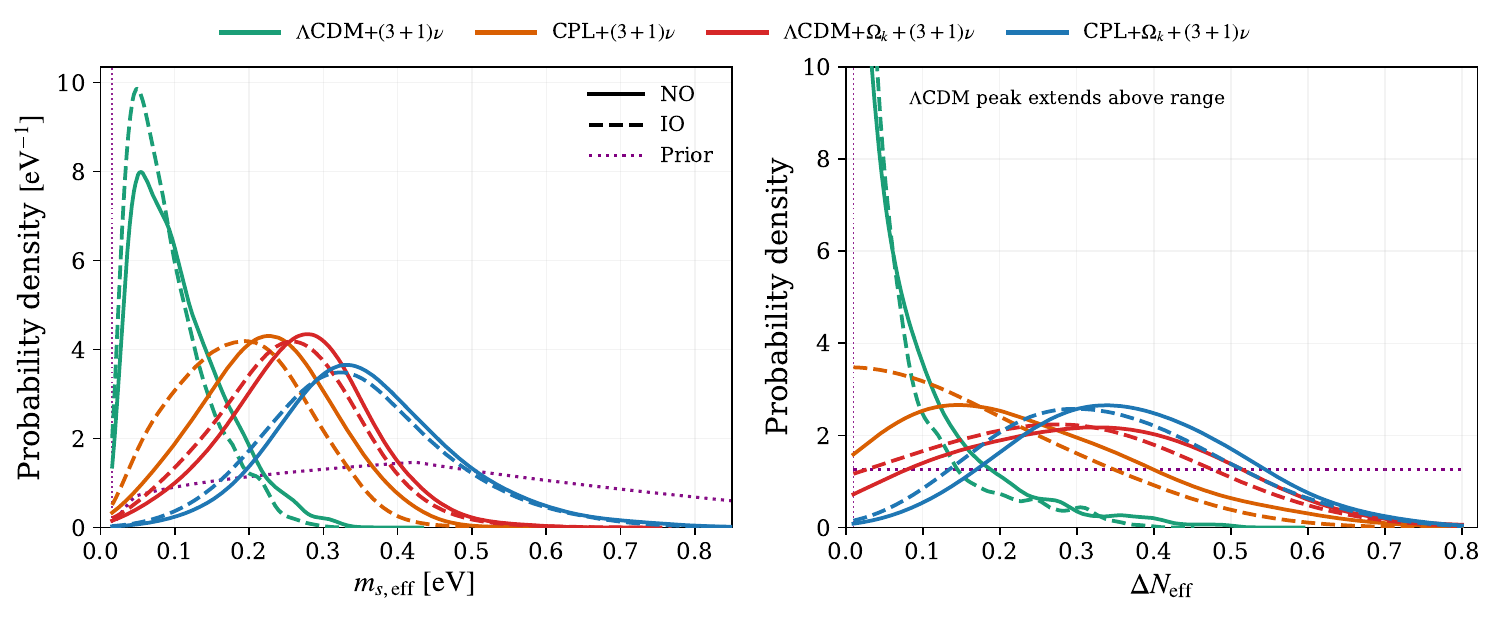}
\caption{ Marginalized posterior densities of the effective sterile-neutrino mass $m_{s,\rm eff}$ \textbf{(left)} and the extra relativistic energy density $\Delta N_{\rm eff}$ \textbf{(right)} for the different cosmological models.
Colors distinguish the cosmological model, while solid and dashed curves denote normal and inverted mass ordering, respectively. The dotted curves show the corresponding priors, and all densities are normalized to unit integral. The vertical dotted lines indicate the absolute lower boundary of the prior.}
\label{fig:m4eff_prior_post}
\end{figure*}

Since the adopted uniform priors on $m_s$ and $\Delta N_{\rm eff}$ induce a non-uniform prior on the derived quantity $m_{s,\rm eff}$, we
compare the unit-normalized marginal posterior densities with the corresponding priors in Fig.~\ref{fig:m4eff_prior_post}. The lower
$2.5$th percentile of the induced $m_{s,\rm eff}$ prior, defined from Eq.~\eqref{eq:induced_prior}, is $q_{0.025}^{\rm prior}(m_{s,\rm eff})\simeq0.064~{\rm eV}$. For the CPL+$\Omega_k$+$(3+1)\nu$ model, the corresponding posterior $2.5$th percentiles are $0.144~{\rm eV}$ and $0.126~{\rm eV}$ for NO
and IO, respectively. Moreover, $99.7\%$ ($99.6\%$) of the posterior probability lies above $q_{0.025}^{\rm prior}(m_{s,\rm eff})$ for NO
(IO). This indicates that the displacement of the posterior toward larger effective sterile masses is not solely a consequence of the
transformed prior volume.
A similar behavior is found for $\Delta N_{\rm eff}$. Its uniform prior has a lower $2.5$th percentile $q_{0.025}^{\rm prior}(\Delta N_{\rm eff})=0.0298$, whereas the corresponding posterior percentiles for the CPL+$\Omega_k$+$(3+1)\nu$ model are $0.102$ and $0.082$ for NO and IO,
respectively. These prior--posterior comparisons quantify the displacement of the sterile-sector constraints away from the low-prior region. They should not, however, be interpreted as a formal exclusion of a vanishing sterile contribution, since the adopted priors impose hard lower boundaries,
$m_{s,\rm eff}^{\rm min}\simeq0.0158~{\rm eV}$ and $\Delta N_{\rm eff}^{\rm min}=0.01$, by construction.

\begin{figure}[t]
\centering
\includegraphics[width=0.9\columnwidth]{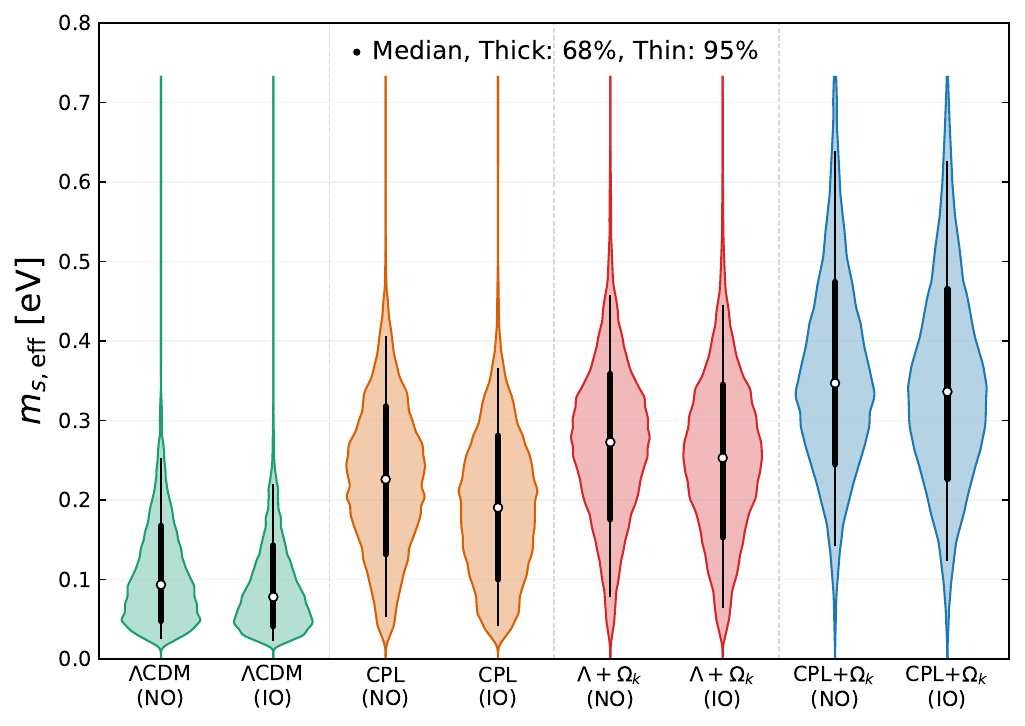}
\caption{ Marginalized posterior distributions of the effective sterile-neutrino
mass $m_{s,\rm eff}$ for the four cosmological model classes. The distributions show a systematic migration toward larger $m_{s,\rm eff}$ when dynamical dark energy and spatial curvature are introduced, with the
largest shift occurring in the CPL+$\Omega_k$+$(3+1)\nu$ models. The violin widths represent the normalized marginal posterior densities, markers and horizontal intervals indicate the posterior median and the $68\%$ and $95\%$ credible intervals, respectively.}
\label{fig:m4eff_violin}
\end{figure}

The same trend is summarized more directly in Fig.~\ref{fig:m4eff_violin}, which shows the marginalized
$m_{s,\rm eff}$ distributions for the four cosmological model classes. For both NO and IO, the posterior shifts systematically toward larger effective sterile masses as CPL dynamics and spatial curvature are introduced.  The largest displacement occurs for the CPL+$\Omega_k$+$(3+1)\nu$ models, consistent with the lower limits
identified from the equal-tail posterior intervals and with the prior-posterior comparison of Fig.~\ref{fig:m4eff_prior_post}.

\begin{figure*}[t]
    \centering
    \includegraphics[width=\linewidth]{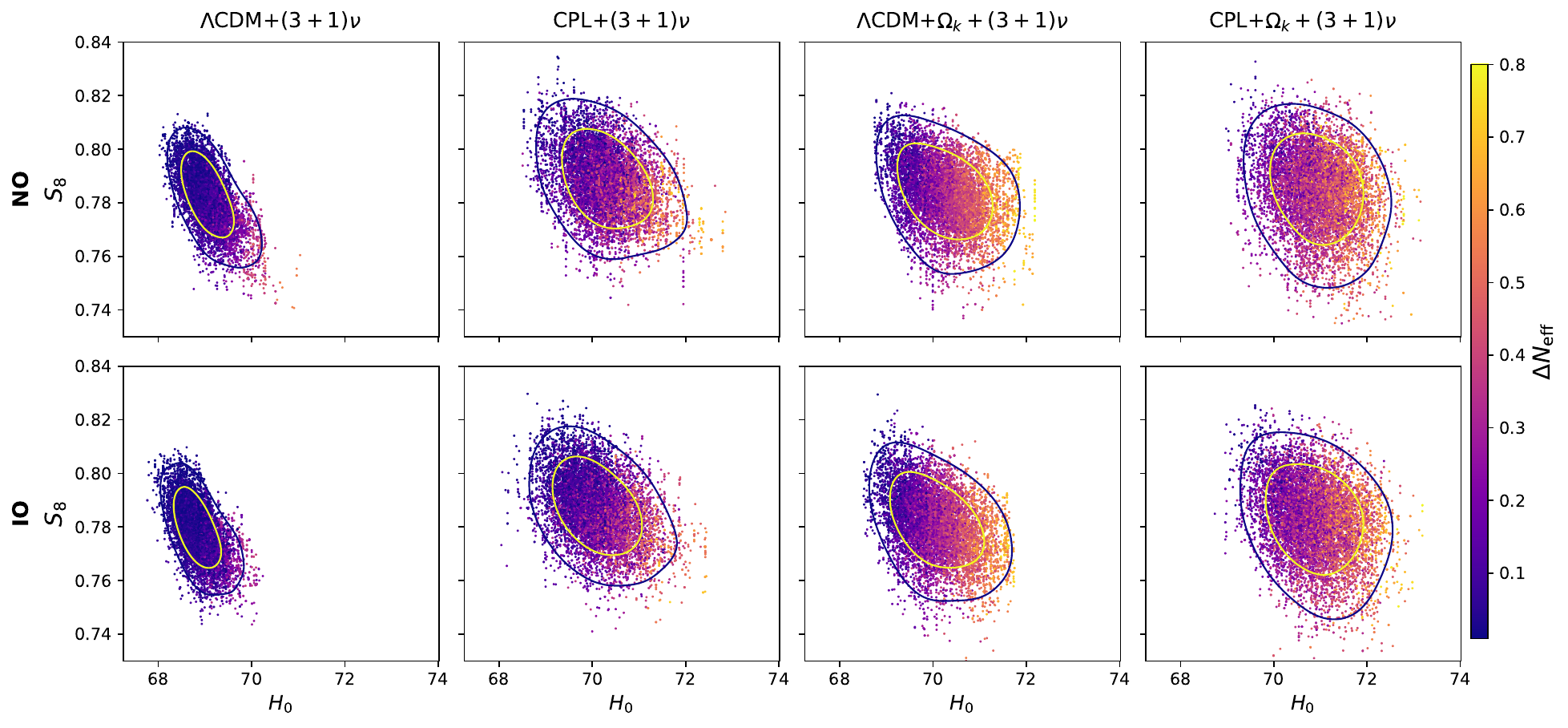}
\caption{
Posterior distributions in the $H_0$--$S_8$ plane for the four
cosmological models considered in this work, obtained with the fiducial
$\mathcal{D}_{\rm PP}$ data combination. The upper and lower rows
correspond to normal inverted mass ordering, respectively,
while the columns show the different late-time cosmological backgrounds.
Posterior samples are colour-coded by $\Delta N_{\rm eff}$ using a common
colour normalization across all panels, allowing the sterile-neutrino
abundance to be compared directly between models and mass orderings.
The solid contours denote the marginalized $68\%$ and $95\%$ credible
regions.}
\label{H0-S8-Ok-dN}
\end{figure*}

A complementary view is shown in Fig.~\ref{H0-S8-Ok-dN}, where the $H_0$-$S_8$ posterior samples for all four cosmological models are colour-coded by $\Delta N_{\rm eff}$. The same colour normalization is used in all panels, allowing the sterile-neutrino abundance to be compared directly across models and mass orderings. The colour gradient reveals a clear correlation between $H_0$ and $\Delta N_{\rm eff}$, with larger $\Delta N_{\rm eff}$ generally associated with larger $H_0$. Allowing dynamical dark energy and spatial curvature also broadens and shifts the joint $H_0$-$S_8$ posterior toward larger sterile-sector abundances, while the overall behaviour remains similar for normal and inverted
mass ordering.

An important feature of the extended models is their behavior in the
$H_0$-$S_8$ plane. Allowing additional freedom in the late-time
background tends to shift the inferred expansion rate toward larger
values, while the clustering amplitude remains close to
$S_8\simeq0.78$--$0.79$. In particular, the CPL+$\Omega_k$+$(3+1)\nu$ model gives
$H_0=70.93\pm0.65~{\rm km\,s^{-1}\,Mpc^{-1}}$ and $S_8=0.785^{+0.015}_{-0.012}$ for NO, with a very similar result for IO. The inferred $S_8$ is consistent with the DES Y6 constraint, $S_8=0.789\pm0.012$, which is not included in our likelihood. Thus, the upward shift in $H_0$ occurs without a comparably large shift in $S_8$. Since the H0DN prior is included in the fiducial data
combination, this should not be interpreted as an independent resolution
of the Hubble tension. Rather, it shows that the extended models can
accommodate the adopted high-$H_0$ information while retaining a
clustering amplitude consistent with the independent DES Y6 determination.

Overall, this shows that spatial curvature,
dark-energy dynamics, and the sterile-neutrino contribution cannot be
interpreted independently within the present parameter space. Allowing CPL evolution redistributes part of the late-time background freedom between $\Omega_k$ and the sterile sector, whereas the active mass ordering has only a secondary effect on these correlations. Whether these additional degrees of freedom are statistically
supported by the data is examined in the following subsection.
\subsection{Relative performance of the cosmological models}
\label{subsec:model_performance}

The relative statistical performance of the models is summarized in
Table~\ref{tab:model_comparison}.  To avoid confusing the effects of the active-neutrino mass ordering with those of the cosmological extensions, each model is compared with the corresponding flat $\Lambda$CDM+$(3+1)\nu$ reference model having the same ordering. Thus, the NO and IO sectors are assessed independently.

\begin{table}[t]
\caption{Model-comparison statistics relative to the corresponding flat $\Lambda$CDM+$(3+1)\nu$ reference model.  The NO extensions are
compared with the NO baseline, and the IO extensions with the IO baseline.  Negative $\Delta\chi^2_{\rm min}$, $\Delta{\rm AIC}$,
and $\Delta{\rm BIC}$ favor the extended model in the corresponding criterion, while positive $\ln B$ favors the extension relative to
its ordering-matched reference.}
\label{tab:model_comparison}
\footnotesize
\centering

\begin{ruledtabular}
\begin{tabular}{lllll}
Model
& $\Delta\chi^2_{\rm min}$
& $\Delta{\rm AIC}$
& $\Delta{\rm BIC}$
& $\ln B$
\\
\hline

\multicolumn{5}{c}{Normal ordering (NO)} \\
\hline

CPL+$ (3+1)\nu$
& $-13.58$
& $-9.58$
& $+3.12$
& $+1.0$
\\

CPL+$\Omega_k$+$ (3+1)\nu$
& $-21.10$
& $-15.10$
& $+3.95$
& $+4.9$
\\

$\Lambda$CDM+$\Omega_k$+$ (3+1)\nu$
& $-13.70$
& $-11.70$
& $-5.35$
& $+6.1$
\\

\hline
\multicolumn{5}{c}{Inverted ordering (IO)} \\
\hline

CPL+$ (3+1)\nu$
& $-15.80$
& $-11.80$
& $+0.90$
& $+3.0$
\\

CPL+$\Omega_k$+$ (3+1)\nu$
& $-24.76$
& $-18.76$
& $+0.29$
& $+7.6$
\\

$\Lambda$CDM+$\Omega_k$+$ (3+1)\nu$
& $-16.20$
& $-14.20$
& $-7.85$
& $+8.2$
\\

\end{tabular}
\end{ruledtabular}
\end{table}

All six extensions improve the maximum likelihood relative to their
ordering-matched reference models.  For the flat CPL+$(3+1)\nu$
models, we obtain
$\Delta\chi^2_{\rm min}=-13.58$ for NO and $-15.80$ for IO.
After accounting for the two additional CPL parameters, these
correspond to $\Delta{\rm AIC}=-9.58$ and $-11.80$, respectively.  Allowing both
CPL dynamics and spatial curvature produces the largest
maximum-likelihood improvement, $\Delta\chi^2_{\rm min}=-21.10\quad({\rm NO}),\,
-24.76\quad({\rm IO}),$ with corresponding $\Delta{\rm AIC}=-15.10$ and $-18.76$.
Hence, according to the AIC, the CPL+$\Omega_k$+$(3+1)\nu$ models provide the strongest improvement among the extensions considered.

The curvature-only $\Lambda$CDM+$\Omega_k$+$(3+1)\nu$ models also provide a substantial
improvement, with $\Delta\chi^2_{\rm min}=-13.70$ and $-16.20$ for NO and IO,
respectively.  Since only one additional cosmological parameter,
$\Omega_k$, is introduced relative to the corresponding reference
model, the AIC penalties are smaller, giving $\Delta{\rm AIC}=-11.70$ for NO and $-14.20$ for IO. The BIC emphasises this economy of parameters even more strongly,
the curvature-only models give $\Delta{\rm BIC}=-5.35$ and $-7.85$, whereas the CPL and
CPL+$\Omega_k$ extensions retain slightly positive BIC differences.
Thus, the information criteria distinguish between the largest
improvement in best-fit likelihood and the most economical extension.

The Bayesian evidence gives a consistent but more global assessment.
With the convention
$\ln B=\ln(\mathcal Z_i/\mathcal Z_{\rm ref})$, positive values favour
the extension relative to its ordering-matched $\Lambda$CDM+$(3+1)\nu$ reference.  We find $\ln B=1.0$ and $3.0$ for the flat CPL models under NO and IO, respectively.  Including curvature increases the evidence to $\ln B=4.9$ for NO and $7.6$ for IO in the
CPL+$\Omega_k$ models.  The largest evidence ratios are obtained for the simpler $\Lambda$CDM+$\Omega_k$ extensions, with $\ln B=6.1$ for NO and $8.2$ for IO.  The Bayesian comparison therefore indicates that much of the statistical improvement can be
captured by introducing spatial curvature alone, without requiring the additional CPL degrees of freedom.

The different ordering of the models under AIC, BIC, and Bayesian evidence reflects the distinct manner in which these quantities penalize model complexity.  The CPL+$\Omega_k$ models achieve the lowest $\chi^2_{\rm min}$ and the most negative $\Delta{\rm AIC}$,
showing that the combined dark-energy and curvature freedom provides the best maximum-likelihood fit.  In contrast, both the BIC and the Bayesian evidence favor the more economical $\Lambda$CDM+$\Omega_k$ extension, which achieves a substantial fit
improvement with only one additional background parameter.  The evidence values should nevertheless be interpreted conditional on the adopted parameter priors.

The improvement in fit quality should also be understood as a property of the joint data combination rather than as an independent preference of every constituent likelihood.  The additional degrees of freedom allow the combined early- and late-time constraints to be accommodated more flexibly, while individual data sets need not separately favor
the corresponding extension.

As a nested-model cross-check, we evaluate the Savage-Dickney density
ratio for the flat limit, $\Omega_k=0$, of the CPL+$\Omega_k$+$(3+1)\nu$ model, as described in Sec.~\ref{subsec:modelcomparison_method}. For ease of interpretation, we quote the inverse Bayes factor, $B_{\rm nonflat,flat}^{\rm SD}
\equiv (B_{\rm flat,nonflat}^{\rm SD})^{-1}$, obtaining $B_{\rm nonflat,flat}^{\rm SD}\simeq2.6\quad ({\rm NO}),\qquad B_{\rm nonflat,flat}^{\rm SD}\simeq6.8\quad ({\rm IO}).$ Both estimates remain stable when the one-dimensional density-smoothing scale is varied over $0.3\leq\texttt{smooth\_scale\_1D}\leq0.6$.
Thus, within the CPL+$ (3+1)\nu$ sector, the nested-model comparison
favors the non-flat realization for both mass orderings, with stronger
support for IO.

Similarly, for the dynamical dark-energy extension we obtain
$
B_{\rm CPL,\Lambda}^{\rm SD}\simeq0.86
\quad ({\rm NO}),\qquad
B_{\rm CPL,\Lambda}^{\rm SD}\simeq1.95
\quad ({\rm IO}),
$
where $B_{\rm CPL,\Lambda}^{\rm SD}>1$ favors the CPL extension over the
cosmological-constant limit. For NO, the Bayes factor is close to unity
and therefore provides little discrimination between the two models.
For IO, the ratio modestly favors the CPL extension. The Savage-Dickey tests indicate that the nested-model support for spatial
curvature is stronger than that for the additional CPL dark-energy
freedom, particularly for IO.

Taken together, the posterior constraints and model-comparison statistics indicate that the main shifts in the inferred cosmology are driven by the combined freedom in the sterile-neutrino, dark-energy, and curvature sectors.  The neutrino mass ordering primarily affects the active-neutrino mass spectrum, while its impact on the late-time cosmological parameters and on the relative statistical performance of the extended models remains comparatively small.  By contrast, allowing spatial curvature and dynamical dark energy produces more appreciable changes in $H_0$, $\Omega_k$, and the sterile-sector parameters, with $S_8$ remaining relatively stable
across the model variants considered.

\section{Summary and Outlook}
\label{sec:conclusions}
We have investigated a partially thermalized sterile-neutrino scenario in a 3+1 framework, allowing simultaneously for dynamical dark energy and spatial curvature. We considered both normal and inverted ordering of the three active neutrinos and combined CMB, BAO, Type-Ia supernovae, redshift-space distortion, and local Hubble-distance measurements. Our results demonstrate that the cosmological inference of sterile-neutrino properties depends strongly on both the sterile-neutrino abundance and the assumed late-time expansion history.

A central result of this work is that allowing partial thermalization permits an eV-scale sterile-neutrino sector to remain compatible with the cosmological data considered here. Rather than assuming a fully thermalized fourth species, we allow its relativistic contribution $\Delta N_{\rm eff}$ and physical mass to vary independently, thereby accounting for a suppressed sterile abundance. We find $\Delta N_{\rm eff}\simeq0.24$ in the flat CPL models and $\simeq$0.34-0.36 in the non-flat CPL models, substantially below the fully thermalized limit $\Delta N_{\rm eff}\simeq1$. Correspondingly, the effective sterile mass increases from $m_{s,\mathrm{eff}}\simeq$0.09-0.11 eV in flat $\Lambda$CDM to 0.19-0.23 eV in flat CPL and $\simeq0.35$ eV in the non-flat CPL models. Thus, the extended cosmologies permit a substantially larger sterile contribution while retaining a subthermal sterile abundance. 

This result is particularly relevant in light of the recent analysis of Ladeira et al., \cite{2026PhRvD.114d3511L} which explicitly examined whether cosmological data exclude sterile neutrinos in general or primarily the fully thermalized scenario. Using Planck, DESI DR2 BAO and supernova data, they found that the fully thermalized sterile scenario is strongly disfavored, whereas partially populated sterile-neutrino realizations remain viable, their different-temperature sterile relic provided the least cosmological tension among the cases considered. They further showed that cosmological data constrain the effective sterile abundance much more strongly than the physical sterile mass \cite{2026PhRvD.114d3511L}. Our results are consistent with this broader picture, but extend it by allowing the sterile abundance to vary continuously within a (3+1) model while simultaneously varying both dark-energy dynamics and spatial curvature. In particular, our non-flat CPL models accommodate $m_{s,\mathrm{eff}}\simeq0.35$ eV with $\Delta N_{\rm eff}\simeq0.35$, illustrating explicitly how a moderately populated sterile species can leave a larger effective mass signature without approaching the fully thermalized limit.

Our results agree with recent DESI DR2 analyses showing that the inclusion of additional radiation can substantially weaken cosmological constraints on sterile-neutrino masses, allowing partially thermalized sub-eV sterile neutrinos to remain viable when combined with a local $H_0$ measurement \cite{2026PhRvD.113b3508E}. Our results also agree qualitatively with other recent DESI analyses. For example, Du et al. found that a $w_0w_a$CDM model with a sterile neutrino can accommodate a nonzero effective sterile mass at approximately the $2\sigma$ level, obtaining $m_{s,\mathrm{eff}}=0.50^{+0.33}_{-0.27}$ eV, whereas their corresponding $\Lambda$CDM and constant-w models did not provide evidence for a massive sterile neutrino \cite{Du:2025iow}. Our inferred values are somewhat lower, $m_{s,\mathrm{eff}}\simeq0.35$ eV in the non-flat CPL models, but exhibit the same qualitative dependence on the freedom in the late-time expansion history. The difference in the numerical constraints is expected given the different datasets, treatment of the sterile distribution, active-neutrino sector, and inclusion of spatial curvature.

Allowing extensions of the late-time cosmology also relaxes the active-neutrino mass constraints. The total active mass increases from $0.0707^{+0.0016}_{-0.0011}$ eV (NO) and $0.1097^{+0.0017}_{-0.0097}$ eV (IO) in flat $\Lambda$CDM to $0.0982^{+0.0059}_{-0.0039}$ eV and $0.1315^{+0.0064}_{-0.0032}$ eV, respectively, in the non-flat CPL models. At the same time, $H_0$ increases from $\sim69$ km/s/Mpc in flat $\Lambda$CDM to $\sim70.9$ km/s/Mpc in the non-flat CPL models. These shifts illustrate the strong degeneracy between neutrino masses, the sterile abundance, and the late-time expansion history. They are also consistent with the broader DESI BAO finding that neutrino-mass constraints can be substantially modified when the assumption of a cosmological constant is relaxed \cite{2026PhRvD.113l3506C, 2025JCAP...02..021A, Ivanov:2026dvl, 2026PDU....5102229B}. Our results further demonstrate that this behavior persists in an enlarged (3+1) neutrino sector, where the active-neutrino mass budget is accompanied by a partially thermalized sterile component. 

The model-comparison statistics further demonstrate the impact of these extensions. Relative to flat $\Lambda$CDM(3+1)$\nu$, the CPL models improve the minimum $\chi^2$ by 13.6-15.8, while the non-flat CPL models improve it by 21.1-24.8. The corresponding AIC differences are negative, $\Delta{\rm AIC}$=-9.6 to -18.8, and the Bayesian evidence increases, with $\ln B=$+1.0 to +8.2. The BIC provides a more conservative assessment, particularly for the CPL models. Thus, while the extended models substantially improve the fit, the statistical evidence depends on the adopted complexity penalty and does not independently establish the necessity of curvature or dynamical dark energy.

The comparison between the normal and inverted active-neutrino orderings reveals a non-negligible dependence of the cosmological fit on the assumed mass hierarchy. In our analysis, the inverted-ordering models consistently yield larger improvements in the minimum $\chi^2$, as well as more favorable AIC values and larger Bayesian evidence ratios, than their normal-ordering counterparts. This trend is seen across both the flat and non-flat dynamical-dark-energy models. 
This hierarchy dependence is noteworthy in light of recent cosmological analyses showing that the inferred neutrino-mass constraints and ordering preference can depend appreciably on the assumed hierarchy, data combination, and cosmological model \cite{2025PhRvD.111h3518H, 2024JCAP...12..020A}. At the same time, recent DESI-based analyses within the standard three-neutrino framework have reported a preference for the normal ordering \cite{Jiang:2024viw, Ivanov:2026dvl}, while global oscillation analyses provide an independent, although not yet overwhelming, sensitivity to the mass ordering \cite{Esteban2024}.
However, our BIC results, provide a more nuanced picture, although the inverted-ordering models have smaller BIC differences than the corresponding normal-ordering models, the positive $\Delta\mathrm{BIC}$ values for the CPL extensions indicate that the additional degrees of freedom are not uniformly favored once the BIC penalty is taken into account. Thus, the apparent preference for the inverted ordering in our extended models should not be interpreted as an independent cosmological determination of the mass hierarchy, but rather as a model-dependent outcome of the combined fit and the adopted model-selection criterion. 

Finally, our results are relevant to the recent discussion of an apparent cosmological preference for an effective neutrino mass below the physical lower bound \cite{2025PhRvD.111h3507G, DESINeutrino2025, 2025PhRvD.111f3534E, 2024PhRvD.110l3537N}. Recent analyses have shown that this preference can be sensitive to CMB lensing and to the assumed late-time cosmology, with dynamical dark energy shifting the inferred effective mass toward physically allowed values \cite{2025PhRvD.111f3534E, 2025PhRvD.112d3541J, 2024JHEP...09..097C,Escudero:2024uea,2026PhRvD.113d3514G}. Our analysis instead keeps all physical neutrino masses non-negative and explores how additional freedom in ($w_0,w_a,\Omega_k$), together with a partially populated sterile sector, modifies the cosmological neutrino constraints. Overall, our results indicate that current cosmological data do not generically exclude the presence of an eV-scale sterile neutrino, but instead place strong constraints on its degree of thermalization and effective cosmological abundance. Future CMB-lensing, BAO, supernova, and large-scale-structure measurements will be crucial for determining whether the present indications for a nonzero sterile contribution persist and for disentangling sterile-neutrino physics from dark-energy dynamics and spatial curvature.

\begin{acknowledgements}
GK acknowledges financial support from Vellore Institute of Technology through its Seed Grant (No. SG20230035, 2023). KP acknowledges financial support from Vellore Institute of Technology through its Seed Grant (No. SG20230039, 2023). The authors also thank the Centre for Technical Support at VIT Vellore for providing the required computational facilities.
\end{acknowledgements}

\appendix

\renewcommand{\thetable}{\Alph{section}.\arabic{table}}
\renewcommand{\thefigure}{\Alph{section}.\arabic{figure}}
\makeatletter
\@addtoreset{table}{section}
\@addtoreset{figure}{section}
\makeatother

\section{Supplementary constraints and robustness tests}

\subsection{Full posterior distributions}
\label{app:triangle}
\begin{figure*}[t]
\centering
\includegraphics[width=0.9\textwidth]{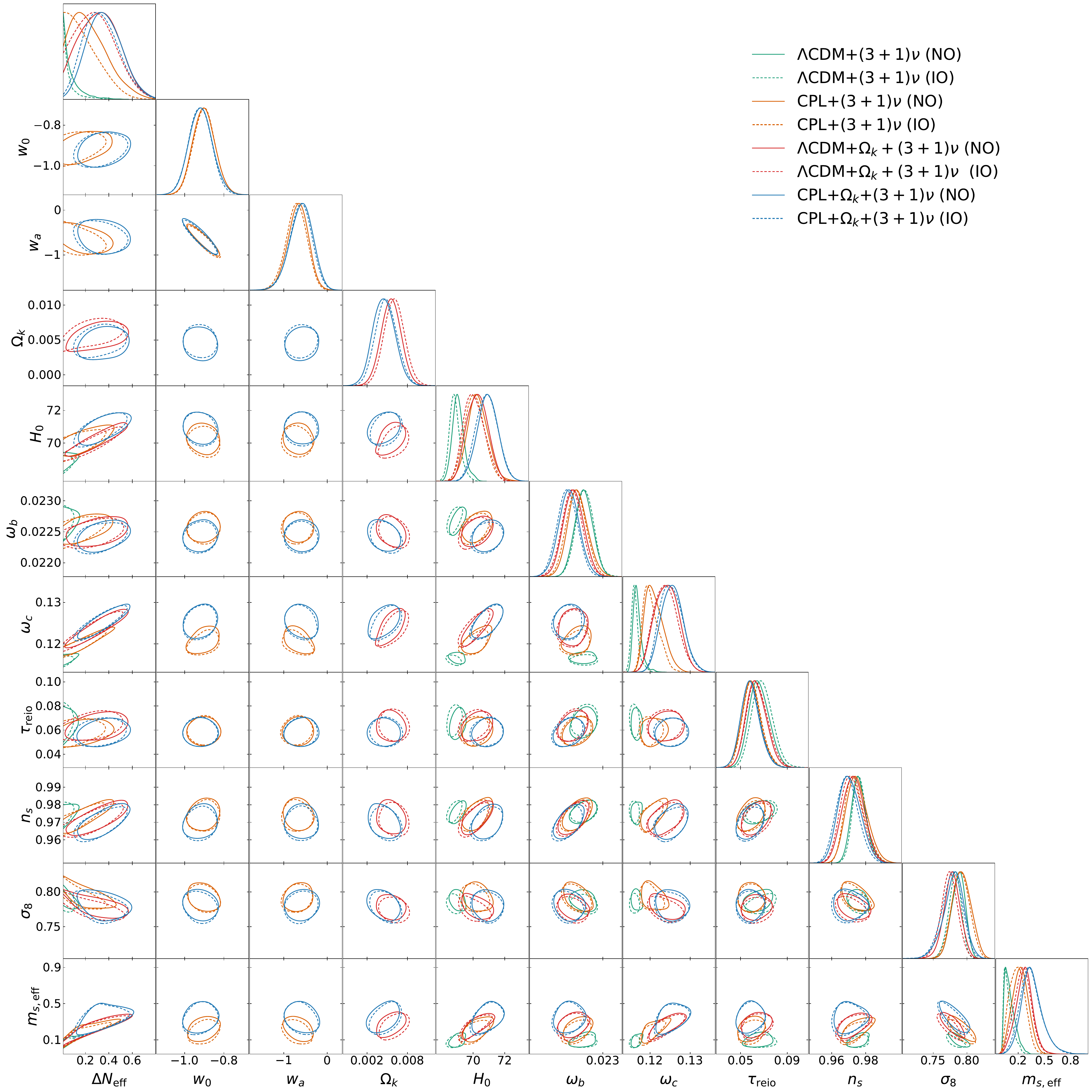}
\caption{ Marginalized one- and two-dimensional posterior distributions of the
    cosmological and sterile-neutrino parameters for the models considered
    in this work using the fiducial Pantheon+ data combination. Contours
    indicate the $68\%$ credible regions.}
\label{fig:all_triangle_PP}
\end{figure*}

Figure~\ref{fig:all_triangle_PP} presents the marginalized one- and
two-dimensional posterior distributions of the main cosmological and
sterile-neutrino parameters for all models using the fiducial
Pantheon+ data combination. The figure provides a compact view of the
parameter correlations and degeneracies underlying the marginalized
constraints discussed in the main text.

\subsection{Baseline cosmological parameters}
\label{app:lcdm_forest}

\begin{figure*}[ht]
\centering
\includegraphics[width=0.9\textwidth]{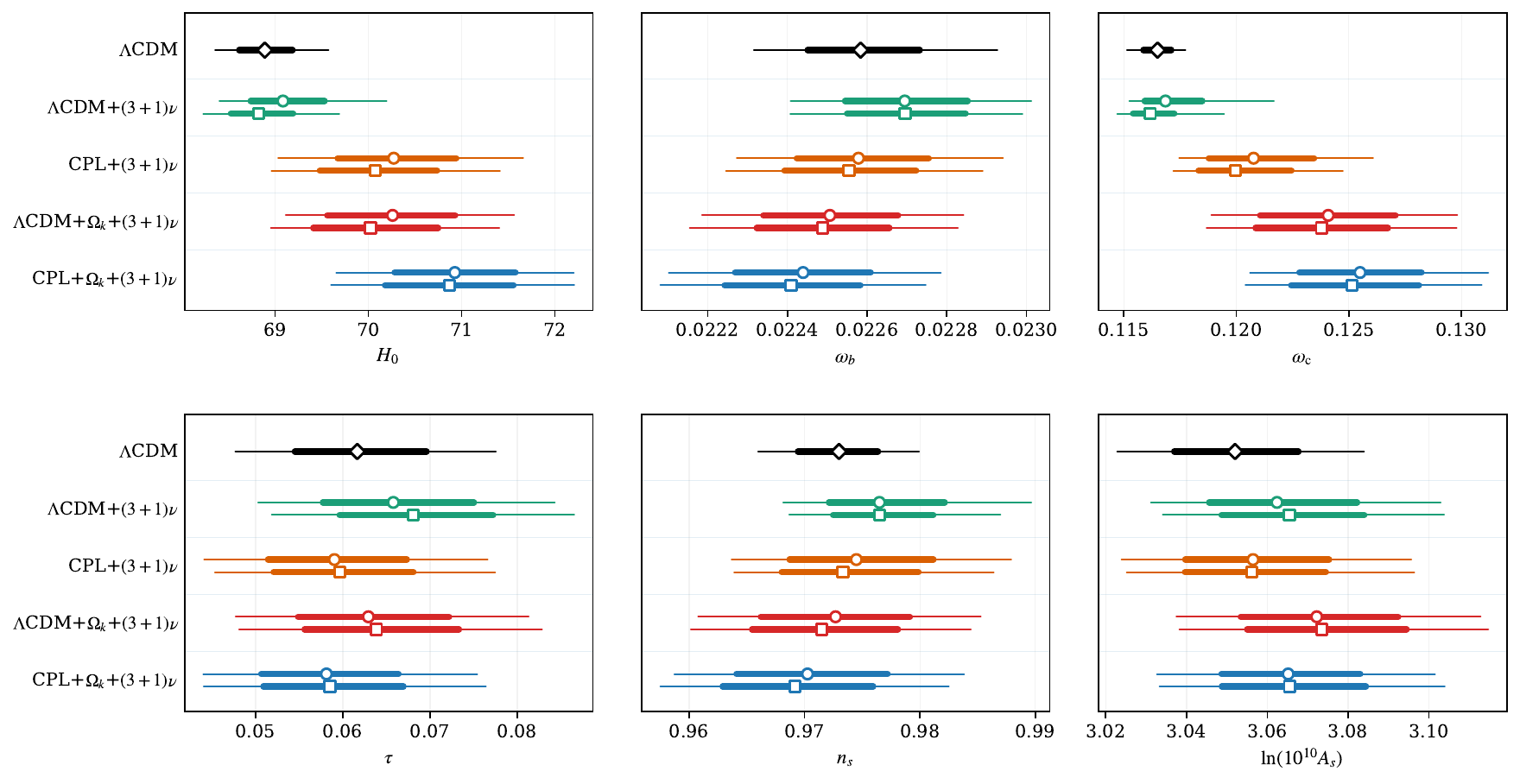}

\caption{ Forest plot showing the marginalized constraints on the six standard cosmological parameters using fiducial Pantheon+ data combination for the different cosmological models considered in this work. Thick and thin horizontal bars denote
the $68\%$ and $95\%$ credible intervals, respectively, while the markers indicate the posterior medians. Circles and squares correspond to normal and inverted ordering, respectively. The standard $\Lambda$CDM result is shown separately as the reference cosmology. The comparison illustrates the shifts and changes in the parameter uncertainties induced by the sterile-neutrino sector, dynamical dark energy, and spatial curvature.}
\label{fig:lcdm6_forest}
\end{figure*}

As a supplementary consistency check, Fig.~\ref{fig:lcdm6_forest}
summarizes the constraints on the six standard $\Lambda$CDM parameters
for the data combinations considered in this work. The forest
representation provides a compact comparison of the posterior central
values and uncertainties and serves as a baseline reference for the
extended cosmological models discussed in the main text.

\subsection{Dependence on the SNe Ia compilation}
\label{app:sn_robustness}
To assess the sensitivity of the inferred constraints to the adopted
supernova compilation, we repeat the analysis of the
CPL+$\Omega_k$+$(3+1)\nu$ model using Union3 \cite{Rubin:2023jdq} in place of Pantheon+, while keeping all other likelihood components unchanged. Table~\ref{tab:u3_pp_comparison} compares the resulting marginalized constraints. The overall agreement between the two analyses indicates that the main conclusions for this model are robust to the choice of SNe Ia compilation.


\begin{table*}[ht!]
\caption{ Marginalized constraints on the parameters of the
CPL+$\Omega_k+(3+1)\nu$ model for normal and inverted ordering, using Union3 and Pantheon+ supernova data.}
\label{tab:u3_pp_comparison}
\centering
\footnotesize
\begin{ruledtabular}
\begin{tabular}{lcccc}
Parameter
& \multicolumn{2}{c}{Union3}
& \multicolumn{2}{c}{Pantheon+} \\
\cline{2-3}\cline{4-5}
& NO & IO & NO & IO \\
\hline

$m_{\rm lightest}\,[{\rm eV}]$
& $<0.0232$
& $<0.0249$
& $<0.0234$
& $<0.0252$ \\

$m_s^{\rm phys}\,[{\rm eV}]$
& $<0.895$
& $<0.901$
& $<0.895$
& $<0.914$ \\

$\Delta N_{\rm eff}$
& $0.39^{+0.15}_{-0.19}$
& $0.35^{+0.14}_{-0.18}$
& $0.36^{+0.14}_{-0.16}$
& $0.34^{+0.13}_{-0.17}$ \\

$w_0$
& $-0.908\pm0.082$
& $-0.912\pm0.081$
& $-0.918\pm0.059$
& $-0.920\pm0.058$ \\

$w_a$
& $-0.63^{+0.34}_{-0.30}$
& $-0.65^{+0.33}_{-0.30}$
& $-0.60^{+0.28}_{-0.24}$
& $-0.61\pm0.26$ \\

$\Omega_k$
& $0.0045^{+0.0015}_{-0.0017}$
& $0.0049\pm0.0016$
& $0.0045\pm0.0016$
& $0.0049\pm0.0016$ \\

$H_0\,[{\rm km\,s^{-1}\,Mpc^{-1}}]$
& $70.96\pm0.73$
& $70.90\pm0.68$
& $70.93\pm0.66$
& $70.87\pm0.67$ \\

$\omega_{\rm b}$
& $0.02245\pm0.00017$
& $0.02241\pm0.00017$
& $0.02244\pm0.00017$
& $0.02241\pm0.00017$ \\

$\omega_{\rm c}$
& $0.1258^{+0.0028}_{-0.0033}$
& $0.1256^{+0.0027}_{-0.0031}$
& $0.1255^{+0.0026}_{-0.0029}$
& $0.1253^{+0.0025}_{-0.0030}$ \\

$\tau_{\rm reio}$
& $0.0585^{+0.0075}_{-0.0085}$
& $0.0585^{+0.0071}_{-0.0084}$
& $0.0586^{+0.0070}_{-0.0087}$
& $0.0588^{+0.0071}_{-0.0087}$ \\

$n_s$
& $0.9712^{+0.0063}_{-0.0073}$
& $0.9697^{+0.0060}_{-0.0072}$
& $0.9705^{+0.0060}_{-0.0070}$
& $0.9694^{+0.0060}_{-0.0069}$ \\

$\ln(10^{10}A_s)$
& $3.066\pm0.018$
& $3.066^{+0.016}_{-0.018}$
& $3.066^{+0.016}_{-0.018}$
& $3.066^{+0.016}_{-0.019}$ \\

$\Omega_{\rm m}$
& $0.3044\pm0.0069$
& $0.3045\pm0.0069$
& $0.3038\pm0.0056$
& $0.3041\pm0.0055$ \\

$\sigma_8$
& $0.779^{+0.017}_{-0.014}$
& $0.777^{+0.017}_{-0.014}$
& $0.780^{+0.017}_{-0.013}$
& $0.778^{+0.016}_{-0.013}$ \\

$S_8$
& $0.784^{+0.016}_{-0.013}$
& $0.783^{+0.015}_{-0.012}$
& $0.785^{+0.015}_{-0.012}$
& $0.783^{+0.015}_{-0.012}$ \\

$\sum m_{\nu,\rm active}\,[{\rm eV}]$
& $0.0964^{+0.0084}_{-0.037}$
& $0.1317^{+0.0047}_{-0.032}$
& $0.0987^{+0.0058}_{-0.039}$
& $0.1315^{+0.0061}_{-0.032}$ \\

$m_{s,\rm eff}\,[{\rm eV}]$
& $0.37^{+0.10}_{-0.14}$
& $0.35^{+0.10}_{-0.13}$
& $0.360^{+0.096}_{-0.14}$
& $0.345^{+0.098}_{-0.14}$ \\

\end{tabular}
\end{ruledtabular}
\end{table*}

\clearpage

\bibliographystyle{apsrev4-2-short}
\bibliography{extracted}

\end{document}